\documentclass[prx,amsmath,amssymb, notitlepage, twocolumn,
nofootinbib,
superscriptaddress
]{revtex4-1}

\usepackage{amsmath}
\usepackage{tabularx,graphicx}
\usepackage{epstopdf}
\usepackage{graphicx}
\usepackage{latexsym}
\usepackage{amssymb}
\usepackage{amsmath}
\usepackage{color}
\usepackage{psfrag}
\usepackage{bbm}
\usepackage{bm}
\usepackage{titlesec}
\usepackage{dsfont}
\usepackage{feynmp}
\usepackage{slashed}

\usepackage{color}
\definecolor{darkblue}{rgb}{0.,0.,0.4}
\definecolor{darkred}{rgb}{0.5,0.,0.}
\usepackage[pdftex,colorlinks=true,linkcolor=darkblue,citecolor=blue,urlcolor=darkred]{hyperref}

\newcommand{\mc}[1]{\mathcal{#1}}

\newcommand{\beq}{\begin{eqnarray}}
\newcommand{\eeq}{\end{eqnarray}}

\newcommand{\la}{\langle}
\newcommand{\ra}{\rangle}

\newcommand{\bsp}{\begin{split}}
\newcommand{\esp}{\end{split}}

\newcommand{\hc}{{\rm h.c.}}

\newcommand{\be}{\begin{equation}}
\newcommand{\ee}{\end{equation}}
\newcommand{\bea}{\begin{eqnarray}}
\newcommand{\eea}{\end{eqnarray}}

\newcommand{\moire} {moir{\' e }}

\renewcommand{\vec}[1]{\bm{#1}}

\makeatletter
\def\l@subsection#1#2{}
\def\l@subsubsection#1#2{}
\makeatother

\begin{document}

\title{Origin of Mott insulating behavior and superconductivity in twisted bilayer graphene}
\author{Hoi Chun Po }
\affiliation{Department of Physics, Harvard University,
Cambridge, MA 02138, USA}
\author{Liujun Zou}
\affiliation{Department of Physics, Harvard University,
Cambridge, MA 02138, USA}
\affiliation{Department of Physics, Massachusetts Institute of Technology,
Cambridge, MA 02139, USA}
 \author{Ashvin Vishwanath}
 \affiliation{Department of Physics, Harvard University,
Cambridge, MA 02138, USA}
\author{ T. Senthil}
\affiliation{Department of Physics, Massachusetts Institute of Technology,
Cambridge, MA 02139, USA}
 \date{\today}
\begin{abstract}
A remarkable recent experiment has observed Mott insulator and proximate superconductor phases in  twisted bilayer graphene when electrons partly fill a nearly flat mini-band that arises a `magic' twist angle.  However, the nature of the Mott insulator, origin of superconductivity and an effective low energy model remain to be determined. We propose a Mott insulator with intervalley coherence that spontaneously breaks U(1) valley symmetry, and describe a mechanism that selects this order over the competing magnetically ordered states favored by the Hunds coupling. We also identify symmetry related features of the nearly flat band that are key to understanding the strong correlation physics and  constrain any tight binding description.  First, although the charge density is concentrated on the triangular lattice sites of the moir{\' e } pattern, the Wannier states of the tight-binding model must be centered on different sites which form a honeycomb lattice. Next, spatially localizing electrons derived from the nearly flat band necessarily breaks valley and other symmetries within any mean-field treatment, which is suggestive of a valley-ordered Mott state, and also dictates that additional symmetry breaking is present to remove symmetry-enforced band contacts.  Tight-binding models describing the nearly flat mini-band are derived, which highlight the importance of further neighbor hopping and interactions.    We discuss consequences of this picture for superconducting states obtained on doping the valley ordered Mott insulator. We show how important features of the experimental phenomenology may be explained and suggest a number of further experiments for the future. We also describe a model for correlated states in trilayer graphene heterostructures  and contrast it with the bilayer case.

\end{abstract}
\maketitle

\tableofcontents

\section{Introduction}
Superconductivity occurs proximate to a Mott insulator in a few materials. The most famous are the cuprate high-$T_c$ materials \cite{Lee2006}; others include layered organic materials \cite{Powell2011}, certain fullerene superconductors \cite{Iwasa2003}, and some iron-based superconductors \cite{Song2016}. In these systems, there is a complex and often poorly understood relationship between the Mott insulator and the superconductor, which has spurred tremendous research activity in condensed matter physics in the last 30 years. Very recently, in some remarkable developments, both Mott insulating behavior and proximate superconductivity have been observed in a very different platform: two layers of graphene that are rotated by a small angle relative to each other \cite{Cao2018, Cao2018a}.

Twisted bilayer graphene (TBG) structures have been studied intensely in the last few years \cite{Neto2007, STMNatPhy, Magaud2010, Mele2010, STMPRL, Bistritzer2011, Castro-Neto2012, Crommie2015, Kim2017, LeRoy2018, Ensslin2018, Ramires}. The charge density is concentrated on a \moire pattern which forms (at least approximately) a triangular lattice \cite{STMNatPhy, Magaud2010, STMPRL,  Crommie2015, Kim2017}. The electronic states near each valley of each graphene monolayer hybridize with the corresponding states from the other monolayer.  When the twisting angle is close to certain discrete values known as the magic angles, theoretical calculations show that there are two nearly flat bands (per valley per spin) that form in the middle of the full spectrum that are separated from other bands.\cite{Bistritzer2011} When the carrier density is such that the chemical potential lies within these nearly flat bands,  interaction effects are expected to be enhanced. At a filling of $1/4$ or $3/4$  (denoted $\nu = -2$ and $+2$ respectively with full band filling denoted $\nu = +4$) of these nearly flat bands, Ref. \onlinecite{Cao2018} reported insulating behavior at very low temperatures. At such fillings band insulation is forbidden, which leads naturally to the expectation that these are correlation-driven (Mott) insulators.  Doping the Mott insulator at $1/4$ band filling - with either electrons or holes - reveals superconductivity at low $T$ \cite{Cao2018a}.

A number of other striking observations have been made in Refs.\ \onlinecite{Cao2018, Cao2018a} about both the Mott insulator and the superconductor from transport studies in a magnetic field. The Mott insulation is suppressed through the Zeeman coupling of the magnetic field at a low scale $\approx 5 T$ - roughly the same scale as the activation gap inferred from zero field resistivity.
Quantum oscillations are seen in the hole doped state with a frequency set (in the hole doped side) by the density deviation from the Mott insulator. The degeneracy of the corresponding Landau levels is half of what might be expected from the spin and valley degrees of freedom that characterize electrons in graphene. The superconductivity occurs at temperatures that are high given the low density of charge carriers. Just like in other doped Mott insulators, there is a dome of superconductivity with $T_c$ reaching an ``optimal" value at a finite doping.
The superconductivity is readily suppressed in accessible magnetic fields - both perpendicular and parallel to the plane.

The observation of these classic phenomena in graphene gives new hope for theoretical progress in addressing old questions on Mott physics and its relationship to superconductivity. They also raise a number of questions. What is the nature of the insulators seen at these fractional fillings? How are they related to the observed superconductivity? On the theoretical side, what is an appropriate model that captures the essential physics of this system?

In this paper we make a start on addressing these questions.  The two nearly flat bands for each valley found in the band structure have Dirac crossings at the \moire K points (but not $\Gamma$). We argue that these Dirac crossings are protected by symmetries of the TBG system. We show that this  precludes finding   a real space representation of the nearly flat bands in terms of Wannier orbitals localized at the triangular \moire sites, in contrast to natural expectations. Thus, a suitable real space lattice model is necessarily  different from a correlated  triangular lattice model with two orbitals (corresponding to the two valleys) per site. We instead show that a representation that is faithful to the Dirac crossings is possible on a honeycomb lattice with two orbitals per site, but even this has some subtleties. First, one cannot implement a natural representation of all the important symmetries in the problem, which include spatial symmetries, time-reversal, and a separate conservation of electrons of each valley (which we dub $U_v(1)$). Second, since the charge density is concentrated at the \moire triangular sites (which appear as the centers of the honeycomb plaquettes), the dominant interaction is not an on-site Coulomb repulsion on the honeycomb sites. Rather it is a `cluster charging energy' that favors having a fixed number of electrons in each honeycomb plaquette. This makes this model potentially rather different from more standard Hubbard models with on-site interactions.

Armed with this understanding of the microscopics, we can begin to address the experimental phenomenology. We propose that this system spontaneously breaks the valley $U_v(1)$ symmetry - we call the resulting order as ``Inter-Valley Coherent" (IVC).  We discuss microscopic mechanisms that stabilize  IVC  symmetry breaking. We point out that even when the IVC is fully polarized it cannot, by itself, lead to a fully insulating state, but rather leads to a Dirac semi-metal.  The development of a true insulator needs a further symmetry breaking (or some more exotic mechanism) to gap out the Dirac points. We show that once the valley symmetry is spontaneously broken, the physics at lower energy scales can be straightforwardly formulated in terms of a real space honeycomb lattice tight-binding model with a dominant cluster charging interaction, and other weaker interactions. We outline a number of different possible routes in which a true insulator can be obtained in such a IVC ordered system.  A concrete example  is a state that further breaks $C_3$ rotational symmetry.   We show how doping this specific IVC insulator can explain the phenomenology of the experiments.  We present  a possible pairing mechanism    due to an attractive interaction mediated by Goldstone fluctuations of the IVC phase. We describe and contrast features of other distinct routes by which the IVC state can become a true insulator at $\nu = \pm 2$. We propose a number of future experiments that can distinguish between the different routes through which a IVC can become a true insulator.

In addition, very recently a heterostructure of  ABC-stacked trilayer graphene and boron nitride (TLG/hBN), which also forms a triangular \moire superlattice even at zero twist angle was studied \cite{Wang2018}. This system also features nearly flat bands that are separated from the rest of the spectrum. Correlated Mott insulating states were seen at fractional fillings of the nearly flat band. Unlike the TBG, here the nearly flat band has no Dirac crossing. This makes the details of the two systems potentially rather different. In particular, the nearly flat band of the TLG/hBN  can be modeled in real space as a triangular lattice model with two orbitals per site, supplemented with  interactions.  However, the hopping matrix elements are, in general, complex (but subjected to some symmetry restrictions).   We describe some properties of this model, and suggest that this system offers a good possibility to realize novel kinds of quantum spin-orbital liquid states.

\section{Electronic Structure of Twisted Bilayer Graphene: General Considerations}
\subsection{Setup}

First, to establish notation, let us consider a graphene monolayer, with lattice vectors $\vec A_1$ and $\vec A_2$  (see Appendix \ref{app:GeoDetails} for details). The honeycomb lattice sites are located at $\vec r_{1,\,2} = \frac{1}{2} (\vec A_1 + \vec A_2) \mp \frac{1}{6} (\vec A_1 - \vec A_2)$, where the $-$ and $+$ signs are respectively for the sites labeled by $1$ and $2$.

Now consider the \moire pattern generated in the twisted bilayer problem.
For concreteness, imagine we begin with a  pair of perfectly aligned graphene sheets, and consider twisting the top layer by an angle $\theta$ relative to the bottom one.  Now we have two pairs of reciprocal lattice vectors, the original ones $\vec B_a$ and $\vec B'_a = R_\theta \vec B_a$. Like references \cite{Neto2007, Bistritzer2011} we approximate the \moire superlattice by the relative wavevectors, leading to a periodic structure with reciprocal lattice vectors $\vec b_a = \vec B'_a -\vec B_a = (R_\theta - I)\vec B_a$. For small $\theta$ we can approximate this by $\vec b_a = \theta {\vec \hat{z}} \times \vec B_a$. Thus, the \moire pattern also has triangular lattice symmetry, but it is rotated by $90 ^\circ$ and has a much larger  lattice constant. Note, in this dominant harmonic approximation, questions of commensuration/ incommensuration are avoided since no comparison is made between the other sets of harmonics of the \moire superlattice that are not commensurate to the dominant one.

Let us now briefly review the low energy electronic structure of monolayer graphene to set the notation. Parameterizing $\vec k =   \sum_{j=1,2} g_j \vec B_j$ for $g_j \in (-1/2,1/2]$, the Bloch Hamiltonian for the nearest neighbor hopping model is:
\begin{equation}\begin{split}\label{eq:}
H(\vec g) =& t (e^{- i \frac{2\pi}{3}(g_1-g_2)} + e^{- i \frac{2\pi}{3}(g_1+2 g_2)} + e^{ i \frac{2\pi}{3}(2g_1 + g_2)}
)  \, \sigma^-\\
&~~~ + {\rm h.c.}
\end{split}\end{equation}
Note that, as a general property of our present choice of Fourier transform, the Bloch Hamiltonian is not manifestly periodic in the Brillouin zone. Rather, for any reciprocal lattice vector $\vec B$, we have $ H(\vec k+ \vec B) = \eta_{\vec B} H(\vec k) \eta_{\vec B}^\dagger$, where $\eta_{\vec B} = {\rm diag} ( e^{- i \vec B \cdot \vec r_a})$. One can now pass to a continuum limit near each  Dirac points $\vec K= (2 \vec B_1 - \vec B_2)/3$ and $- \vec K$. We then have the linearized Hamiltonian
\begin{equation}\begin{split}\label{eq:}
H(\vec K + \vec k) = - \hbar v_F\,  \vec k  \cdot \vec \sigma;\\
H(-\vec K + \vec k) = \hbar v_F \,\vec k \cdot \vec \sigma^*,\\
\end{split}\end{equation}
where $\hbar v_F = \sqrt{3} t a/2$ in our simple nearest-neighbor model. Since $H(\vec q)$ is not periodic in the BZ, expanding about the other equivalent Dirac points will lead to a slightly modified form of the Hamiltonian (due to conjugation by some $\eta$). In second quantized notation we can write the continuum Hamiltonian:
\begin{equation}\begin{split}\label{eq:Hk}
 \hat h_{+} =&   - \hbar v_F \int d^{\!\!\!-2} \vec k \,
\hat  \psi_{+;  \vec k}^\dagger   \left(  \vec k \cdot \vec \sigma \right)\hat  \psi_{+;  \vec k};\\
 \hat h_{-} =& + \hbar v_F  \int d^{\!\!\!-2}  \vec k'\,
\hat  \psi_{-;  \vec k'}^\dagger   \left(  \vec k' \cdot \vec \sigma^* \right)\hat  \psi_{-;  \vec k'}.
\end{split}\end{equation}
where the momentum integration is understood to be implemented near the Dirac point momentum by introducing a cutoff $|\vec k| \leq \Lambda$ and $d^{\!\!\!-2} \vec k=\frac{dk_xdk_y}{(2\pi)^2}$.  The symmetry implementation on the continuum fields is tabulated in Appendix \ref{app:GeoDetails}. For example, $C_3$-rotation symmetry is represented as 
$\hat C_3 \hat \psi_{\pm \mu;  \vec k} \hat C_3^{-1} =\,  e^{\mp i \frac{2 \pi }{3}\sigma_3 }\hat \psi_{\pm \mu ;  C_3 \vec k}$, where $\mu = {\rm t} ,{\rm b}$ is the layer index.

Next, we couple the degrees of freedom in the two layers of graphene and arrive at a continuum theory for the twisted bilayer graphene system \cite{Neto2007,Bistritzer2011}.
First, we note that the rotated Bloch Hamiltonian of a monolayer can then be identified as $H_{\varphi}( \vec k) = H(R_{-\varphi} \vec k)$. Linearizing about the rotated K point $R_{\varphi} \vec K$, obtains $\hat h_{\pm}(\varphi) $, with $ \hat h_{\pm}(\varphi)$ defined by replacing $\vec \sigma \mapsto \sigma_{\varphi} $ in $\hat h_{\pm }$, where
\begin{equation}\begin{split}\label{eq:sigTheta}
\vec \sigma_{\varphi} \equiv e^{- i \varphi \sigma_3/2} \,\vec \sigma\,e^{ i \varphi \sigma_3/2}.
\end{split}\end{equation}
Focusing on a single valley, say $\vec K$, The continuum theory \cite{Neto2007,Bistritzer2011} of the twisted bilayer graphene system is described by the Hamiltonian $\hat H_{\rm Cont.} = \hat H_{\rm Dirac} + \hat H_{\rm T}$, where
\begin{equation}\begin{split}\label{eq:CTh}
\hat H_{\rm Dirac}  =&  \hat h_{+}(\varphi_{\rm t}) + \hat h_{+}(\varphi_{\rm b});\\
\hat H_{\rm T} =&  \int_0^{\Lambda} d^{2} \vec k \, \hat  \psi_{+{\rm b};  \vec k}^\dagger \,  T_{\vec q_1}\,  \hat  \psi_{+{\rm t};  \vec k + \vec q_1}\, + {\rm h.c.}\\
&~~~+ \text{symmetry related terms},
\end{split}\end{equation}
${\rm t}$ and ${\rm b}$ respectively denote the top and bottom layers, and we set $\varphi_{\rm t} = \theta/2$ and $\varphi_{\rm b} = - \theta/2$.
Here, we have introduced $\vec q_1 \equiv R_{-\theta/2} \vec K- R_{\theta/2}\vec K$, which characterizes the momentum transfer between the electronic degrees of freedom of the two layers \cite{Bistritzer2011}. 
Assuming $T_{\vec q_1}$ is real, the symmetries of the system, which we will discuss in the following subsection, constrain $T_{\vec q_1}$\cite{Mele2011} to take the form $T_{\vec q_1}= w_0 - w_1 \sigma_1$, where  $w_{0,1}$ are real parameters \footnote{Note that, due to a different choice of basis, our parameterization is slightly different from that of Refs.\ \onlinecite{Bistritzer2011,Mele2011}. We have therefore defined $w_1$ with an additional minus sign to account for the difference in basis choice.}. Similarly, one can generate the omitted symmetry-related terms by applying symmetries on $T_{\vec q_1}$.

\subsection{Symmetries of the Continuum Theory}
Let us discuss how the symmetries  of the graphene monolayer are modified in the twisted bilayer problem within the dominant harmonic approximation. We will see that in addition to the \moire translation symmetry, we have C$_6$ rotation, time reversal and a mirror symmetry. Furthermore, a U(1) valley symmetry that allows us to assign valley charge to the electrons emerges in the low-energy limit. The generator of C$_6$ rotation and time reversal will flip the valley charge, while reflection leaves it invariant.

Microscopically, the stacking pattern of the two layers can be specified as follows \cite{Bistritzer2011,Mele2011,ShiangPRB}: first, we align the two layers perfectly in a site-on-site manner, corresponding to the ``AA stacking'' pattern, and then rotate the top and bottom layers about a hexagon center by angles $\theta/2$ and $-\theta/2$ clockwise respectively; second, we shift the top layer by a vector $\vec d$ parallel to the plane.
For generic values of $\theta$ and $\vec d$ one expects that almost all of the spatial symmetries are broken.

\begin{figure}[h]
\begin{center}
{\includegraphics[width=0.45 \textwidth]{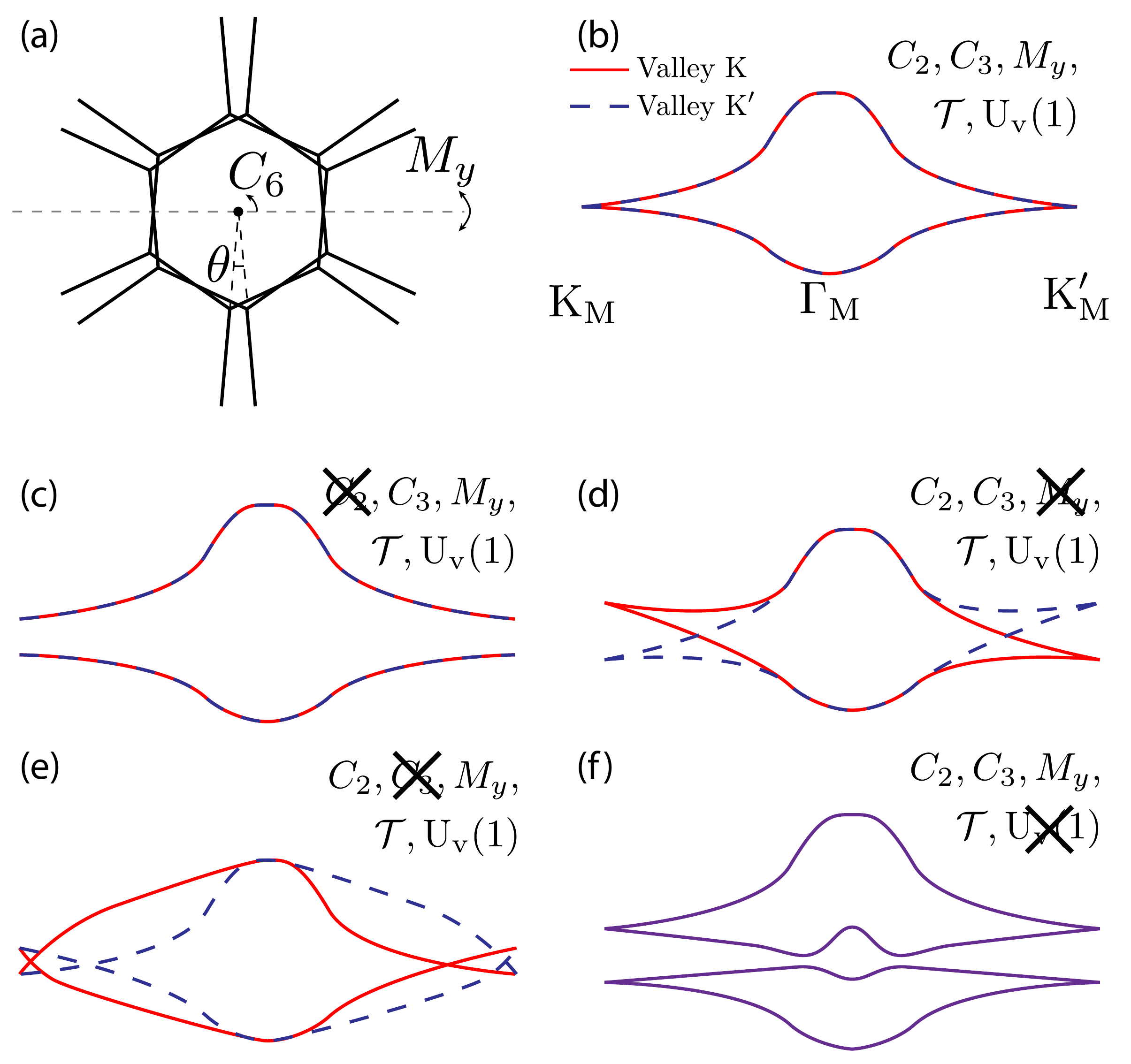}}
\caption{
{\bf Effective symmetries and constraints on band structures.} (a) The effective symmetries of the twisted bilayer graphene system can be inferred by inspecting the point-group symmetry of a hexagon center in the real space, taken to be the rotation axis of the layers. (b) Schematic band structure along a high-symmetry path in the \moire Brillouin zone. (c-f)  Effect of symmetry breaking.
(c) Breaking the $C_2$ rotation will gap out the Dirac points. 
(d) An external perpendicular electric field breaks the mirror $M_y$ symmetry, which only modifies the energetics but cannot open a band gap at charge neutrality \cite{Neto2007}. 
(e) When $C_3$ rotation is broken, but the combined symmetry of two-fold rotation $C_2$ and time-reversal $\mathcal T$ is preserved, the Dirac points remain protected, although unpinned from ${\rm K_M}$ and ${\rm K_M'}$. (f) When valley conservation ${\rm U_v}(1)$ symmetry is broken, one can no longer label the bands using their valley index. 
The gaplessness at charge neutrality is no-longer symmetry-required, although, depending on detailed energticss, there can still be remnant Dirac points.
In contrast, at quarter filling relevant for the observed Mott physics, there are necessarily Dirac points present in this case. The other symmetry breaking patterns listed above also do not open band gaps at quarter filling.
\label{fig:Sym}
 }
\end{center}
\end{figure}

However, within the dominant harmonic approximation it was found that, on top of possessing \moire lattice translation symmetries, the effective theory is also insensitive to $\vec d$ \cite{Bistritzer2011}. This implies that, given $\theta$, the effective theory will at least possess all the exact symmetries for any choice of $\vec d$. A particularly convenient choice is when we take $\vec  d = \vec 0$. In this case, we can infer all the point-group symmetries of the system by focusing on the center of the hexagons (Fig.\ \ref{fig:Sym}a). Aside from the rotational symmetries generated by the six-fold rotation $C_6$, we see that there is an additional mirror plane $M_y$, which, in fact, combines a mirror perpendicular to the 2D plane together with an in-plane mirror which flips the top and bottom layers. Strictly speaking, this leads to a two-fold rotation in 3D space, but when restricting our attention to a 2D system it acts as a mirror.

To summarize, the effective theory will at least have the following spatial symmetries: lattice translations, a six-fold rotation and a mirror. This allows one to uniquely identify its wallpaper group (i.e., 2D space group) as $p6mm$ (numbered 17 in Ref.\ \onlinecite{ITC}). Having identified the symmetries of the system, one can derive the model following a phenomenological approach by systematically incorporating all symmetry-allowed terms with some cutoff \cite{Mele2011}. We have tabulated the explicit symmetry transformation of the electron operators in Appendix \ref{app:GeoDetails}.

In the effective theory the degrees of freedom arising form the microscopic {\rm K} and {\rm K'} points are also essentially decoupled \cite{Neto2007, Bistritzer2011}. This is because, for a small twist angle satisfying $|\sin \theta| \ll 1$, we have $| \vec b_a| \ll | \vec K|$, and therefore the coupling between the {\rm K} and {\rm K'} points is a very high-order process. Hence, on top of the usual electron-charge conservation, the effective theory has an additional, emergent ${\rm U_v}(1)$ conservation corresponding to the independent conservation of charge in the two valleys {\rm K} and {\rm K'}. Henceforth, we will refer to this as  ``valley conservation.'' The valley charge operator is given by:

\begin{equation}\label{eq:IzDef}
\hat I_z = \int d^{\!\!\!-2} \vec k \,\left (
\hat  \psi_{+;  \vec k}^\dagger   \hat  \psi_{+;  \vec k} - \hat  \psi_{-;  \vec k}^\dagger   \hat  \psi_{-;  \vec k} \right )
\end{equation}
Note that, as time-reversal $\mathcal T$  interchanges the {\rm K} and {\rm K'} valleys, it is not a symmetry of a single valley. Similarly, $C_6$ also interchanges the two valleys, thus:
\beq
\hat {\mathcal T} \hat I_z \hat {\mathcal T}^{-1} &=& - \hat I_z\\
\hat {C_6} \hat I_z \hat {C_6}^{-1} &=& - \hat I_z\\
\hat M_y \hat  I_z \hat {M_y}^{-1} &=& +\hat I_z
\eeq
We then see that their combined symmetry, $C_6 \mathcal T$, is a symmetry in the single-valley problem. In fact, one can check that the symmetries of single-valley problem is described by the magnetic space group 183.188 (BNS notation; Ref.\ \onlinecite{IUC}). We have tabulated the generating symmetries in Table \ref{tab:Sym}.

\begin{center}
\begin{table}
\caption{{\bf Summary of key effective symmetries.} From top to bottom, the listed symmetries are time-reversal, \moire lattice translation, a perpendicular 2D mirror, three-fold rotation, combined symmetry of two-fold rotation and time-reversal, and valley ${\rm U_v}(1)$ conservation. For any symmetry $g$, it either commutes ($\eta_g = +1$) or anticommutes ($\eta_g = -1$) with the valley charge operator $\hat I_z$.
\label{tab:Sym}}
\begin{tabular}{c|c|c}
Symmetry & $\eta_g$ & Remarks\\
\hline
$\mathcal T$ & $-1$ & Broken by  valley polarization $\langle I_z\rangle \neq 0$\\
$t_{\vec a} $ & $+1$ & --\\
$M_y $ & $+1$ & Broken by perpendicular electric field\\
$C_3 $ & $+1$ & Pins Dirac points to ${\rm K_M}$ and ${\rm K_M'}$  \\
$C_2 \mathcal T $ & $+1$ & ~Protects the local stability of the Dirac points~\\
$\exp{(- i \theta \hat I_z)}$ ~&$+1$ & --
\end{tabular}
\end{table}
\end{center}

\section{Low Energy Theory: Two Band Projection}

Formally, the continuum effective theory \cite{Neto2007,Bistritzer2011} we described corresponds to an infinite-band problem for each valley. However, near charge neutrality it was found that, for some range of angles, the \moire potential can induce additional band gaps at certain commensurate filling of the \moire unit cell. The ``nearly flat bands'' identified near the magic angle correspond to two bands per valley, separated from all other bands by band gaps, that form Dirac points at the ${\rm K_M}$ and ${\rm K_M'}$ points in the \moire BZ.
These bands correspond to the relevant degrees of freedom for the correlated states observed in Refs.\ \onlinecite{Cao2018,Cao2018a}, and in the following we will focus our attention to the properties of these bands. In this section, we will always focus on a single valley, say that corresponding to the K point in the microscopic description.

\subsection{Symmetry-enforced Band Contacts \label{sec:BandContants}}
A salient feature of the effective theory is the presence of Dirac points at charge neutrality, whose velocity is strongly renormalized and approaches zero near the magic angle \cite{Neto2007, Bistritzer2011}. The stability of the Dirac points can be understood from symmetries: for a single valley, ${\rm K_M}$ is symmetric under the (magnetic) point-group generated by $C_6 \mathcal T$. In particular, $(C_6 \mathcal T)^2 = C_3$, and therefore we  can label each band at ${\rm K_M}$ by its $C_3$ eigenvalue, which takes value in $\{ 1, \omega = e^{-i \frac{2\pi}{3}}, \omega^* \}$. In particular, a band with $C_3$ eigenvalue $\omega$ is necessarily degenerate with another with eigenvalue $\omega^*$, as $(C_6 \mathcal T)^2 \neq 1$ on these bands and enforces a Kramers-like degeneracy. The observed Dirac points at charge neutrality correspond precisely to this two-dimensional representation (Fig.\ \ref{fig:Sym}b).

While we have alluded to the presence of $C_6 \mathcal T$ symmetry in explaining the stability of the Dirac points, these band contacts are actually locally stable so long as the symmetry $ (C_6 \mathcal T)^3 = C_2 \mathcal T$ is kept. This can be reasoned by noting that $ C_2 \mathcal T$ quantizes the Berry phase along any closed loop to $0, \pi  \mod 2\pi$, and a Dirac point corresponds precisely to the case of a nontrivial $\pi$ Berry phase \cite{C2TInvariant}.

Let us now consider the effect of breaking the various symmetries (spontaneously or explicitly) in the system. First, as $C_2 \mathcal T$ is crucial in protecting the local stability of the Dirac points, once it is broken  the Dirac points can be immediately gapped out (Fig.\ \ref{fig:Sym}c).
However, as long as $C_2 \mathcal T$ symmetry is preserved, a small breaking of any other point-group symmetries will not lead to a gapped band structure at charge neutrality. For instance, the mirror $M_y$ maps K$_{\rm M}$ to K$_{\rm M}$', and its presence only ensures that the two inequivalent Dirac points are at the same energy. Therefore, even when a perpendicular electric field is externally applied such that $M_y$ is broken, as in the setup of Refs.\ \onlinecite{Cao2018,Cao2018a}, it can only induce an energy difference between the two Dirac points \cite{Neto2007} (Fig.\ \ref{fig:Sym}d).
This should be contrasted with the case of Bernal-stacked bilayer graphene, whose quadratic band touching at charge neutrality can be gapped by an external electric field \footnote{This can be understood by noting that, in the case of Bernal stacking, the system has an effective $C_2$ symmetry which interchanges the top and bottom layer. A perpendicular electric field will therefore break this symmetry, and hence the band degeneracy at charge neutrality is lifted.}.
Alternatively, if $C_3$ symmetry is broken the Dirac points are unpinned from ${\rm K_M}$ and ${\rm K_M'}$ (Fig.\ \ref{fig:Sym}e). As such, for a sufficiently strong $C_3$ breaking,
a band gap might open at charge neutrality if the Dirac points could meet their oppositely charged partners and annihilate. (Though, as we will argue later, this is impossible without further symmetry breaking.)
\cite{Goerbig2017}.

Now consider the case when valley conservation is spontaneously broken by an IVC, i.e., the valley charge $\hat I_z$ is no longer conserved. In this case, we should first consider the full four-band problem consisting of both valleys. At, say, K$_{\rm m}$, the combined symmetry of $M_y \mathcal T$ ensures that the Dirac points from the two valleys are degenerate. While such degeneracy is lifted in the presence of an IVC, as long as the remaining symmetries are all intact we can only split the degeneracy according to $4 = 2 \oplus 2$ (Fig.\ \ref{fig:Sym}f). This remaining two-fold degeneracy rules out an interpretation of the experimentally observed Mott insulator as a Slater insulator with a spatial-symmetry-respecting (ferro) IVC incorporated at the Hartree-Fock level. Instead, one must either introduce additional symmetry breaking, say that of $C_3$ or lattice translations, or consider an IVC which also breaks some additional spatial symmetries.  We will elaborate on these points in Sec.\ \ref{sec:IVC-Gen}. We also note that, an essentially identical argument holds for the case of spontaneously ferromagnetic order leading to fully spin-polarized bands of $I_z$ ordering. In this way it connects to the quarter-filled Mott insulator we will be interested in.

\subsection{Triangular versus Honeycomb Lattice}

A conventional route for understanding the correlated states observed in Refs.\ \onlinecite{Cao2018,Cao2018a} is to first build a real-space tight-binding model for the relevant bands, and then incorporate short-range interactions to arrive at, say, a Fermi-Hubbard model. Typically, the orbital degrees of freedom involved in the tight-binding model can be identified from either applying chemistry insight, or more systematically by studying the projected density of states for the relevant bands, both of which are inapplicable to the current \moire potential problem; furthermore, understanding on the structure of the wave-functions is required. Indeed, as is noted in Refs.\ \onlinecite{Magaud2010, PabloPRL, ShiangPRB}, the local density states for the flat bands are well-localized to the AA regions of the \moire pattern, which form a triangular lattice.
This theoretical prediction has also been confirmed experimentally \cite{STMPRL, STMNatPhy,Crommie2015}. Based on this observation, it is natural to consider a real-space model starting from effective orbitals centered at the AA sites, which corresponds to a tight-binding model defined on the triangular lattice \cite{Cao2018, Cao2018a}. In addition, by treating the two valleys separately, one envisions a model with two orbitals localized to each of the triangular sites (i.e., AA regions of the \moire pattern).

From symmetry representations, however, we can immediately rule out such a model. This can be readily inferred from the computed band structure \cite{Neto2007,Bistritzer2011,Mele2011,ShiangPRB} (Fig.\ \ref{fig:WFBands}j): While the two bands are nondegenerate at $\Gamma$, as we have explained they form symmetry-protected Dirac points at K$_{\rm M}$ and K$_{\rm M}$'. Using such pattern of degeneracies one can infer the possible symmetry representations at these high-symmetry points, and from a real-space analysis \cite{NC, TopoChem, MSG} one finds that a triangular-lattice model will always leads to the same symmetry representation at all three of the high-symmetry points, i.e., they are either all nondegenerate, or are all Dirac points. This is inconsistent with the observed pattern of degeneracies, which rules out all triangular-lattice models.

In fact, the degeneracy pattern described is familiar---it corresponds exactly to the monolayer graphene problem. One can further check that this is the only possible solution using the methods described in Refs.\ \onlinecite{NC,MSG}.
Symmetry-wise, this implies that any tight-binding model must correspond to orbitals forming a honeycomb lattice. To reconcile with the predicted and observed local density of states \cite{Magaud2010, ShiangPRB,STMPRL, STMNatPhy,Crommie2015}, however, these orbitals must have nontrivial shapes: although each orbital is centered at a honeycomb site, which corresponds to the AB/BA region of the \moire pattern, the weight of the orbitals are mainly localized to the AA sites. Therefore, we expect the shape of the orbitals to resemble a (three-lobed) fidget spinner (Figs.\ \ref{fig:WF}(a,b)).

\subsection{Obstructions to Symmetric Wannier States \label{sec:Obstructions}}

Our symmetry analysis suggests that one should model the system by orbitals centered at the AB/BA regions of the \moire potential, which form a honeycomb lattice. A minimal tight-binding model of a single valley would then be
\begin{equation}\begin{split}\label{eq:}
\hat H_{\rm Minimal} =   \sum_{\vec \rho_i} \, t_{\vec \rho_i} e^{i \phi_{\vec \rho_i} }\hat c^\dagger_{\vec r} \hat c_{\vec r + \vec \rho_i}  + {\rm h.c.},
\end{split}\end{equation}
where $\hat c_{\vec r}^\dagger$ is an electron creation operator centered at a honeycomb site (for a single valley), and $\vec \rho_i$ connects two $i$-th nearest neighbor sites. Given that this describes a single valley which breaks time reversal symmetry, the hoppings are in general complex unless constrained by a space-group symmetry.

A pedestrian approach would involve optimizing the parameters $\{  t_{\vec \rho_i}, \phi_{\vec \rho_i}\}$ to reproduce the energy eigenvalues obtained from the continuum description. Would this be a good starting point for building up a real-space effective model upon which we can incorporate interaction terms? Contrary to usual expectations, we will argue that such an approach  has a serious flaw in capturing certain essential properties.
Specifically, we will show that while the energy eigenvalues may be well-approximated, the topology of the resulting Bloch wave-functions will necessarily be incorrect. This has important dynamical consequences, relating to the stability of band contacts under different symmetry assumptions, which in turn dictate whether an insulator will result at particular fillings. In particular, we found two symmetry obstructions to deriving a single-valley tight-binding model.
The first concerns the symmetry representations of $M_y$: we found that the two bands have opposite $M_y$ eigenvalues of $\pm 1$, whereas, from a real-space analysis \cite{NC,TopoChem,MSG}, one can show that the two bands in a tight-binding model must have the same $M_y$ eigenvalue.

There is a second, more serious, obstruction: aside from a quantized Berry phase of $\pi$ for any closed loop encircling a single Dirac point, one can further define a $\mathbb Z$-valued winding number \cite{Goerbig2017}. In contrast to the conventional case of graphene, the two inequivalent Dirac points in the single-valley model are known to have the \emph{same} winding number \cite{CastroNeto2011, Goerbig2017,Cao2018}. As the net winding number of the Dirac points arising in any two-band tight-binding model would necessarily be zero, we can then conclude that there is an obstruction for a symmetric real-space description, i.e., there is an obstruction for constructing localized Wannier functions that reproduces just the two bands of interest, represents $C_2 \mathcal T$ naturally, and preserves valley quantum numbers. A more detailed description of this obstruction, by relating it to the anomalous surface state of a three dimensional topological phase, is contained in the Appendix \ref{app:classAIII}. Essentially, this argument invokes three key ingredients: (i) a two band model and (ii) $C_2{\mathcal T}$ symmetry and (iii) net winding of the Dirac points  in the Brillouin zone.

We will return to the question of tight binding models in Section \ref{sec:TB}, but for the discussion below, we will work directly in the momentum space in the manifold of states spanned by the nearly flat bands.

\section{ Inter-valley coherent order: phenomenological motivation}

We first describe some important clues from experiments\cite{Cao2018,Cao2018a} on the nature of both the Mott state and the superconductor.  We  begin with the observation that - at optimal doping - an in-plane magnetic field suppresses the superconductivity when the Zeeman energy scale is of order the zero field $T_c$. This shows that the superconductor has spin-singlet pairing.  Upon hole doping the $\nu = -2$ insulator, quantum oscillations are seen with a frequency set by the density of doped holes in perpendicular $B$-fields exceeding $\approx 1 T$. This tells us that the ``normal" metallic state and the superconductor that emerges from it should be regarded as  doped Mott insulators: the charge carriers that are available to form the normal state Fermi surface or the superconducting condensate are the doped holes. Thus  the hole-doped superconductor retains information about the Mott insulator.  In contrast, electron doping this Mott insulator leads very quickly to quantum oscillations with a high frequency that is set by the deviation of charge density from the charge neutrality point ($\nu = 0$). This may indicate a first order transition between a metal and Mott insulator on the electron doped side. It will be important to search for signs of hysteresis in transport experiments as the gate voltage is tuned. As the superconductor is better developed and characterized on the hole doped side, we will restrict attention to hole doping from now on.

A further important clue from the quantum oscillation data is that the Landau levels (per flux quantum) is two-fold degenerate, whereas one would expect four-fold degeneracy coming  from the spin and valley degeneracy. The doped holes have thus lost either their spin or valley quantum numbers (or some combination thereof). Losing spin makes it hard to reconcile with spin singlet pairing that can be suppressed with a Zeeman field.  Thus, we propose instead that the valley quantum number is lost.  The simplest option\footnote{A more exotic option to explain the reduced Landau level degeneracy should also be kept in mind. Instead of losing the valley quantum number by symmetry breaking we lose it through fractionalization. For instance the electron could split into a fermion that carries its charge and spin but not the valley quantum number and a charge-$0$, spinless boson that carries its valley quantum number. If the boson is gapped while the fermion forms a fermi surface in the doped state we will get the reduced Landau level degeneracy. Of course such fractionalization will come hand in hand with an emergent gauge field. } then is that the valley quantum number is frozen due to symmetry breaking, {\em i.e} $\langle \vec I \rangle \neq 0$.
Here, we may define $\vec I$ using the electron operators $\hat c (\vec k)$ for the nearly flat band states:
\begin{equation}\begin{split}\label{eq:}
\vec I = \sum_{a,b,n, \alpha, \vec k}  \hat c_{a n \alpha }(\vec k)^\dagger \vec \tau_{ab} \, \hat c_{b n \alpha }(\vec k),
\end{split}\end{equation}
where $a,b = \pm $ correspond to the valley index, $\alpha$ is the spin index,  $n$ labels the two bands for each valley, and $\vec \tau$ denotes the standard Pauli matrices.

A non-zero expectation value for $ I^z$ breaks time reversal symmetry. This will lead to a sharp finite temperature phase transition in $2d$, and would likely have been detected in the experiments. Given the absence of any evidence of a sharp finite temperature transition we propose that the ordering is in the pseudospin $xy$ plane. These phenomenological considerations therefore lead us to a IVC ordered state.

We note that, for IVC ordering to be useful to explain the quantum oscillations, it has to occur at a scale that is large compared to the scales set by the magnetic field.  Specifically, the band splitting due to IVC ordering must be bigger than the Landau level spacing  $\approx 15-30 K$ at the biggest fields used (of order $5 T$). This means that the IVC order is much more robust than the superconductivity and occurs at a higher temperature scale. We further need the IVC order to be present already in the Mott insulator, so that upon doping it can impact the quantum oscillations.

Thus, our view is that the first thing that happens as the sample is cooled from high temperature is IVC ordering.  This order then sets the stage for other phenomena to occur at lower temperature (the Mott insulation, or the superconductivity).

\section{Simple theory of the IVC ordered state}

We now describe a mechanism that stabilizes  IVC ordering, and describe the properties of the resultant state. Interestingly, to treat this stage of the problem it is sufficient to work within a momentum space formulation. This enables us to sidestep the difficulties elaborated in Sec.\ \ref{sec:Obstructions} with a real space tight-binding formulation.

Consider the nearly flat bands in the limit of strong Coulomb repulsion. Note that the dominant part of the interaction is fully $SU(4)$ invariant. We expect that the Coulomb interaction prefers an $SU(4)$ ferromagnetic state - similar  to the $SU(4)$ ferromagnetism favored by Coulomb interaction in the zeroth Landau level in monolayer graphene\cite{Nomura2006,Alicea2006,Young2012}, or in the  extensive literature on  flat band ferromagnetism\cite{Tasaki1998}.  Indeed, the difficulties with Wannier localization of the nearly flat bands also suggest that, when Coulomb interactions dominate, an $SU(4)$ ferromagnetic ground state will be favored. The band dispersion, however, is not $SU(4)$ symmetric, and hence there will be a selection of a particular direction of polarization in the $SU(4)$ space.  To address this, we consider the energies of different orientations of the $SU(4)$ ferromagnet within a simple  Hartree-Fock theory. Specifically we compare a spin polarized state, a pseudospin $I^z$ polarized state, and the IVC state with $I^x$ polarization.

Assume a Hamiltonian
\beq
\label{HFHam}
H = H_0 + V
\eeq
with
\beq
\label{NFband}
H_0=\sum_{an\alpha\vec k}\epsilon_{an}(\vec k)c_{an\alpha}^\dag(\vec k) c_{an\alpha}(\vec k)
\eeq
Similarly to before, $a$ is the valley index, $\alpha$ is the spin index, and $n$ labels the two bands for each valley. The dispersion $\epsilon_{an}(\vec k)$ is independent of the spin, and due to time reversal $\epsilon_{an}(\vec k)=\epsilon_{-an}(-\vec k)$.  We assume a simple form of interaction:
\beq
\label{HFint}
\begin{split}
V&=\frac{g}{N}\sum_{\vec k_1\vec k_2\vec q}c^\dag_{an\alpha}(\vec k_1+\vec q)c_{an\alpha}(\vec k_1)\\
&\quad\quad\quad\quad\quad
\cdot c^\dag_{a'n'\alpha'}(\vec k_2-\vec q)c_{a'n'\alpha'}(\vec k_2)
\end{split}
\eeq
where $N$ is the number of $\vec k$-points in the \moire Brillouin zone. Repeated indices are summed over here. This interaction actually has an $SU(8)$ symmetry, but this is strongly broken down to $SU(4)$ by the difference in dispersion between the two bands, and eventually down to $U(2) \times U(2)$ by the asymmetry of the dispersion under $k \rightarrow -k$.  Each $U(2)$ factor corresponds simply to independent $U(1)$ charge and $SU(2)$ spin conservation symmetries of the two valleys.

We also remark that Eq.\ \eqref{HFint} is overly simplified, for it does not incorporate form factors arising from the modulation of the Bloch wave-functions over the BZ when projecting onto the nearly flat bands. With such form factors included, the interaction projected onto the nearly flat bands should be written as
\beq
\begin{split}
	V&=\frac{g}{N}\sum_{\vec k_1\vec k_2\vec q}\Lambda^{a}_{nn'}(\vec k_1+\vec q, \vec k_1)\Lambda^{a'}_{mm'}(\vec k_2-\vec q, \vec k_2)\\
	&\cdot c^\dag_{an\alpha}(\vec k_1+\vec q)c_{an'\alpha}(\vec k_1)
	\cdot c^\dag_{a'm\alpha'}(\vec k_2-\vec q)c_{a'm'\alpha'}(\vec k_2)
\end{split}
\eeq
with the form factors given by the Bloch wave functions of the states in the nearly flat bands via
\beq
\Lambda^a_{nn'}(\vec k_1, \vec k_2)=\la u_{an}(\vec k_1)|u_{an'}(\vec k_2)\ra
\eeq
where $|u_{an}(\vec k)\ra$ is the Bloch wave function of a state in the nearly flat bands labelled by valley index $a$, band index $n$ and momentum $\vec k$ (it has no dependence on the spin indices). These form factors are potentially important for the present problem due to the nontrivial band topology present in the valley-resolved band structure. Our preliminary analysis suggests that the results of the Hatree-Fock calculation are modified in the ultra-flat-band limit, i.e., when the interaction term overwhelms the kinetic energy, whereas the key conclusions below are stable within a range of intermediate interaction strengths. 
In view of this, in the following we will first pursue the simplified Hatree-Fock theory, and leave the task of settling down the real ground state for future (numerical) studies; it is an interesting question answering if the actual experimental systems demand a more sophisticated treatment.

Details of the Hartree-Fock calculation are presented in Appendix \ref{app:HF}. To summarize, we find that the IVC state has lower energy than both spin and $I^z$ polarized states. The physical reason is that, for both the spin and $I^z$ polarized states, the order parameter is conserved and hence there is a linear shift of the band when the order parameter is non-zero. In contrast, due to the $k \rightarrow -k$ dispersion anisotropy, the IVC order parameter does not commute with the Hamiltonian. IVC order thus does not simply shift the band but  modifies it more significantly.  Assuming a near full polarization in the Hartree-Fock Hamiltonian, the non-commutativity leads to an extra energy gain at second order in the IVC state compared to the spin polarized or $I^z$ polarized states.

Note that, in the presence of $U(2) \times U(2)$ symmetry, the spin singlet IVC state is degenerate with states that have  spin triplet IVC ordering with an order parameter $I^x \vec S$.
The selection between the singlet and triplet IVC order has to occur due to other terms in the Hamiltonian that have been ignored so far.  We will not attempt to pin down the details of this selection in this paper and will simply assume, as suggested by the phenomenology, that the spin singlet IVC is preferred, and discuss its consequences.

Next, we turn to a description of the properties of the IVC state.  We assume that the order parameter is large, and first study its effects on the band structure.  In the absence of valley ordering, at the two Dirac points there is a four-fold band degeneracy.  As explained in Section \ref{sec:BandContants}. The valley ordering splits this four-fold degeneracy into two sets of two-fold degenerate Dirac points. When the order parameter is large, the four nearly flat bands split into two sets of two bands (Fig.\ \ref{fig:Sym}f). At quarter filling, we fill the bottom most band. This, however, results in not a Mott insulator but a Dirac semi-metal. Thus, the IVC state by itself does not lead to a Mott insulator, and a further mechanism is needed. We discuss this in the next section. We note that the semimetals obtained from planar valley order versus $I_z$ order are rather different, the latter being similar to spin ordered states. Furthermore, while additionally breaking $C_3$ symmetry alone can eventually gap the Dirac points of the IVC semimetal, the same is not true of spin or $I_z$ ordered semimetals, which need further symmetry breaking due to their Dirac points carrying the same chirality.

Going beyond the mean field,  the  universal properties of the IVC ordered state are determined by its symmetry breaking. It will have a Goldstone mode with linear dispersion at the longest scales. Further, it will have a finite temperature BKT transition which will have weak signatures in standard experimental probes\footnote{In principle, as discussed earlier the $U_v(1)$ symmetry is only approximate and will be weakly broken by, say, small corrections to the band structures or by 3-body interaction terms. This will modify the physics very close to the phase transition.}

\section{ Intervalley coherent Mott insulators: generalities and a  concrete example \label{sec:IVC-Gen}}
\begin{figure}[h]
\begin{center}
{\includegraphics[width=0.4 \textwidth]{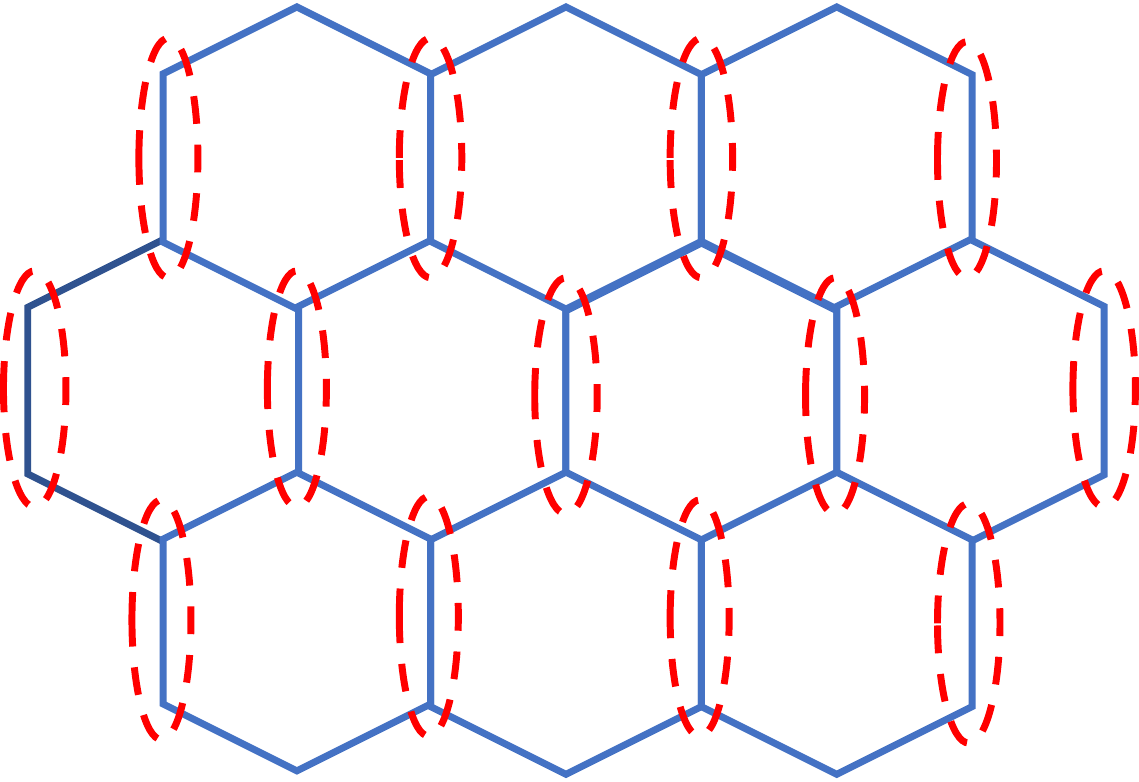}}
\caption{{\bf A $C_3$ symmetry breaking order which preserves other symmetries.}
\label{fig:C3-Break}
 }
\end{center}
\end{figure}

We saw that IVC ordering by itself only gives us a Dirac semi-metal and not a Mott insulator. We now consider the physics below the IVC ordering scale. First, we note that that once $U_v(1)$ symmetry is broken, there is no difficulty with writing down a real space tight-binding model for the two lowest bands. This model lives on the honeycomb lattice and must be supplemented with interactions.  The dominant interaction will be the cluster charging energy  penalizing charge fluctuations on each honeycomb plaquette. Thus, a suitable model Hamiltonian at scales much smaller than the intervalley coherence scale takes the form
\bea
H & = & H_t + H_U \\
H_t & = & -\sum_{ rr'} \sum_\alpha  t_{rr'}  c^\dagger_{r\alpha } c_{r'\alpha} + h.c \\
H_U & = & U\sum_R (Q_R - 2)^2
\eea
where $r, r'$ are sites of the honeycomb lattice, $R$ is the center of the honeycomb plaquettes ({\em i.e} the triangular \moire sites) and we have introduced the cluster charge $Q_R$ defined through
\be
Q_R = \sum_{r \in hexagon} \sum_{r\alpha} \frac{n_{\alpha}(r) }{3}
\end{equation}
We have also specialized to  $\nu = -2$ when this honeycomb lattice is half-filled. For the usual Hubbard model with a strong  on-site repulsion, the Mott insulating state has the usual 2-sublattice Neel order. However, when the cluster charging energy is dominant, this is not obviously the case. We will therefore allow ourselves to consider a few different possibilities for the Mott insulator.   Naturally, in all these options the charge gap of the insulator will be much lower than the scale of IVC ordering. In the experiments, the charge gap is estimated to be about $5 K$. The IVC ordering should then occur at a much higher scale, consistent with what we already concluded based on the phenomenology.  In this section, to be concrete,  we focus on a particular Mott insulator where the $C_3$ rotation symmetry is spontaneously broken while preserving other symmetries (Fig.\ \ref{fig:C3-Break}).

\subsection{$C_3$ broken insulator}

Breaking the $C_3$ symmetry allows  gapping  out the Dirac points and leads to an insulator.   As the $C_3$ breaking order parameter increases, the two Dirac points will move towards each other (Fig.\ \ref{fig:Sym}e) and eventually annihilate to produce a fully gapped insulator.  This annihilation (and correspondingly the gap minimum just into the insulator) will occur either at the $\Gamma$ or $M$ point depending on details of the dispersion.
Note that, within this picture, the $C_3$ breaking also occurs at a scale bigger than the  $\approx 5 K$ charge gap of the Mott insulator.  Clearly, the excitations above the charge gap are ordinary electrons, and their gap can be readily closed by a Zeeman field.

Upon doping this insulator, charge will enter as ordinary holes and form a small Fermi pocket. This pocket will be centered at either $\Gamma$ or $M$ depending on the location of the minimum insulating gap.  In either choice, due to the absence of $C_3$ symmetry, there will just be a single such Fermi pocket which will accommodate the full density of doped holes. Due to the intervalley  ordering these holes will be valley polarized in the $I^x$ direction.  Naturally this explains the quantum oscillation experiments---the frequency will be set by the density of doped holes, and the Landau level degeneracy (per flux quantum) will only be two-fold (from the spin).

A natural pairing mechanism emerges from the coupling of the holes to Goldstone fluctuations of the intervalley order, as we now elaborate.
In the presence of an intervalley condensate an appropriate effective action will be
\begin{eqnarray}
S & = & S_0[\psi]  + S_1 [\psi, \theta] \\
S_0 & = &  \int d\tau \left( \int d^2x \bar{\psi} \left( \partial_\tau + \mu \right) \psi + \int d^2k ~ \bar{\psi}_k h_k \psi_k  \right) \\
S_1 & = & \int d\tau d^2x ~ \Phi_0 (e^{- i\theta} \bar{\psi}_+ \psi_-  + c.c)
\end{eqnarray}
Here $\psi$ is a continuum electron field that represents the electrons in the low-energy nearly flat bands, $\theta$ is the phase of the intervalley condensate and $\Phi_0$ is its amplitude. Note that $h_k = \epsilon_s(k) + \epsilon_a(k) \tau^z$ is a $2 \times 2$ matrix for each $k$ point\footnote{Strictly speaking we should allow for all 4 bands per spin and work with a four coponent $\psi$ and a corresponding $4 \times 4$ Hamiltonian. However for the present discussion we will eventually only be interested in the modes in the vicinity of the Fermi surface after the flavor ordering. It is this sufficient to focus on the two lower bands that are split off by the flavor ordering. We therefore focus on just these two right from the start.}.  We will allow for slow Goldstone fluctuations of the phase and obtain a convenient form of the electron-electron interaction induced by these fluctuations. To that end, we first define new fermion variables $\chi$ through
\be
\psi = e^{\frac{i\theta \tau_z }{2}} \chi.
\ee
This removes the $\theta$ dependence from $S_1$, but $S_0$ now takes the form
\bea
S_0[\psi] & = & S_0[\chi] + S_0'[\chi, \theta] \\
S_0'[\chi, \theta] & = & \int_{x, \tau}  \frac{i}{2} \partial_\tau \theta \bar{\chi} \tau^z \chi +  \frac{1}{2} \partial_i \theta J^v_i(x)
\eea
Here $J_v$ is the contribution to the $U_v(1)$ current from the fermions. It is conveniently written down in momentum space as
\be
J^v_i(q) = \int d^2 k~ \bar{\chi}_{k+q} \frac{\partial h_k}{\partial k_i} \chi_k
\ee
Now we assume that $\Phi_0$ is near maximum polarization and diagonalize the $\chi$ Hamiltonian obtained from $S_0[\chi] + S_1[\chi]$. As discussed earlier, there are two sets of bands per spin (corresponding to $I_x = \pm 1$) that are well separated from each other. The low energy electrons are those that have valley polarization $I_x \approx 1$. We wish to obtain the coupling of these electrons to the $\theta$ fluctuations. For the bands with $I_x \approx 1$, we write
\be
\chi_{+\alpha} = \chi_{-\alpha} \equiv d_\alpha
\ee
It follows that $\bar{\chi} \tau^z \chi \approx 0$ and similarly $\bar{\chi} \frac{\partial \epsilon_s(k)}{\partial k_i} \tau^z \chi_k \approx 0$.  The only non-vanishing coupling therefore is to the contribution from $\epsilon_a(k)$. We get
\be
J^v_i(q) \approx \int d^2 k ~  \bar{d}_{k+q} \frac{\partial \epsilon_a(k)}{\partial k_i} d_k \equiv  \int d^2 k ~  v^a_i (k) \bar{d}_{k+q}   d_k
\ee

Now we assume we have integrated out the fermions everywhere except in the close vicinity of the Fermi surface. This gives a long wavelength, low frequency effective action for the $\theta$ fluctuations of the form
\be
S_{eff}[\theta] = \int_{q, \omega}K \left(\frac{\omega^2}{v^2} + q^2 \right) |\theta (q, \omega)|^2
\ee
Here $K$ is the phase stiffness of the $\theta$ field, and $v$ is the velocity of the linear dispersing $\theta$ fluctuations. We now integrate out $\theta$ to get an effective interaction between the $c$-electrons:
\be
S_{int} =-  \int_{q, \omega}  \frac{q^2}{32 K (\frac{\omega^2}{v^2} + q^2)} |J^v_i (q, \omega)|^2
\ee
This is an attractive interaction. Anticipating that the important regime for pairing is $|\omega| \ll vq$ for an approximate treatment we set $\omega = 0$ in the prefactor to get a simplified effective interaction
\be
S_{int} = - \frac{1}{32 K} \int_{q, \omega} |J^v_i(q, \omega)|^2
\ee
We emphasize again that - within our approximate treatment - the only contribution to $J^v_i$ comes from the {\em antisymmetric} part of the ``normal" state dispersion.
This attractive interaction can now be treated within BCS mean field, and will lead to a superconducting state.

Note that, in real space, since the large repulsion will be on the hexagon center and not on the honeycomb site, there is no particular reason to disfavor on-site s-wave pairing.  Though we will not give a detailed description of the pairing symmetry, the route to superconductivity sketched above naturally leads to a spin singlet superconductor. Further, it forms out of a `normal' metal of ordinary holes through a BCS-like pairing mechanism. We expect then that Zeeman fields of order $T_c$ will efficiently suppress the superconductivity except possibly at very low doping (where eventually phase fluctuations will kill $T_c$).  At low doping, and when one is near a high symmetry point of the Brillouin zone (which is consistent with the fact that there is no additional degeneracy seen in quantum oscillations), the antisymmetric part of the dispersion is expected to constrained by symmetry to be small. For example  near the $\Gamma$ point, it vanishes as the cube of the crystal momentum. This would lead to a small valley current (the derivative of the antisymmetric dispersion with respect to momentum) and hence a weakening of the coupling to valley Goldstone modes, as the doping is reduced. However, if $C_3$ rotation symmetry is broken,  the antisymmetric dispersion can include a term that is linear in momentum, leading to a non-vanishing valley current at small doping.

\section{Other possible Mott insulating states}

The $C_3$ broken insulator is a concrete example of how an intervalley condensate of the twisted bilayer system can eventually become a Mott insulator.  However, given the current experimental information, it is not clear that this is uniquely dictated. Therefore, we sketch a few different Mott insulating states and present some of their phenomenological consequences.

\begin{enumerate}

\item
Translation broken insulator: Broken \moire translations -  for instance Kekule ordering on the  effective honeycomb lattice - can also gap out the Dirac points. The properties of this state and its evolution into the doped superconductor will be similar to the $C_3$ broken insulator discussed above.

\item
Antiferromagnetic insulator: This is the familiar Mott insulator of the usual honeycomb Hubbard model. Upon doping it is expected to evolve into a spin-singlet superconductor as seen in numerical studies of the $t-J$ honeycomb model\cite{Gu2013}. The pairing symmetry appears to be $d+id$. It will be interesting to look for signatures of broken time reversal symmetry if this scenario is realized. Further, this state is known to have quantized spin and thermal Hall effects, and associated gapless edge states\cite{Senthil2000,Volovik1997}. Other properties related to this state are also discussed in the literature\cite{Nandkishore2012, Nandkishore2014}.

\item
Featureless Mott insulators: Given that the honeycomb lattice features two sites in the unit cell, it evades the Lieb-Shultz-Mattis theorem and allows for a featureless ground state (i.e. a gapped insulator with neither topological order nor symmetry breaking) at half filling \cite{Kimchi, Ran, Bauer, Metlitski, Jian, JYLee}. Pictorially, this is viewed as a spin singlet Cooper pair of electrons being localized on orbitals composed of equal weight superpositions of the hexagons of the honeycomb lattice. While model wave functions of this phase have been constructed, the interactions that can drive a system into this phase remain to be understood.

\item
Quantum spin liquids: The simplest possibility is a fully gapped quantum spin liquid. In this case there are neutral spin-$1/2$ excitations (spinons) in the insulator.  Upon doping  a natural possibility is that the charge goes in as bosonic holons (spinless charge-$e$ quasiparticles) whose condensation leads to superconductivity. This is the classic RVB mechanism \cite{Lee2006} for superconductivity in a doped Mott insulator. However, in this scenario, at low doping the superconducting $T_c$ will not have anything to do with the spin gap (measured by the Zeeman scale needed to suppress pairing).

\end{enumerate}

We do not attempt to decide between these different options in this paper. However, we will outline  experiments that can distinguish between them in Sec. \ref{sec: experimental proposal}.

\section{Tight Binding Models \label{sec:TB}}
\subsection{For twisted BG}

As we have argued, there is an obstruction for writing down any tight-binding model for the single-valley problem.
A natural way out, therefore, is to instead consider the four-band problem consisting of both valleys. Our earlier argument requiring the orbitals to be centered on the sites of the honeycomb lattice still applies, but now with two orbitals associated with every site. In addition, to resolve the mirror-eigenvalue obstruction we described, these two orbitals should transform oppositely under the mirror symmetry. These orbitals, however, cannot have definite valley charge, for otherwise the problem is reduced back to the earlier case with each valley considered separately. Instead, it is natural to demand each of the two orbitals to be a time-reversal singlet, which would lead to a standard representation for the symmetry group of $p6mm$ together with time-reversal. 
However, due to the aforementioned anomaly the representation of valley ${\rm U_v}(1)$ is necessarily complicated, and we will address that later.

\begin{figure}
\begin{center}
{\includegraphics[width=0.48 \textwidth]{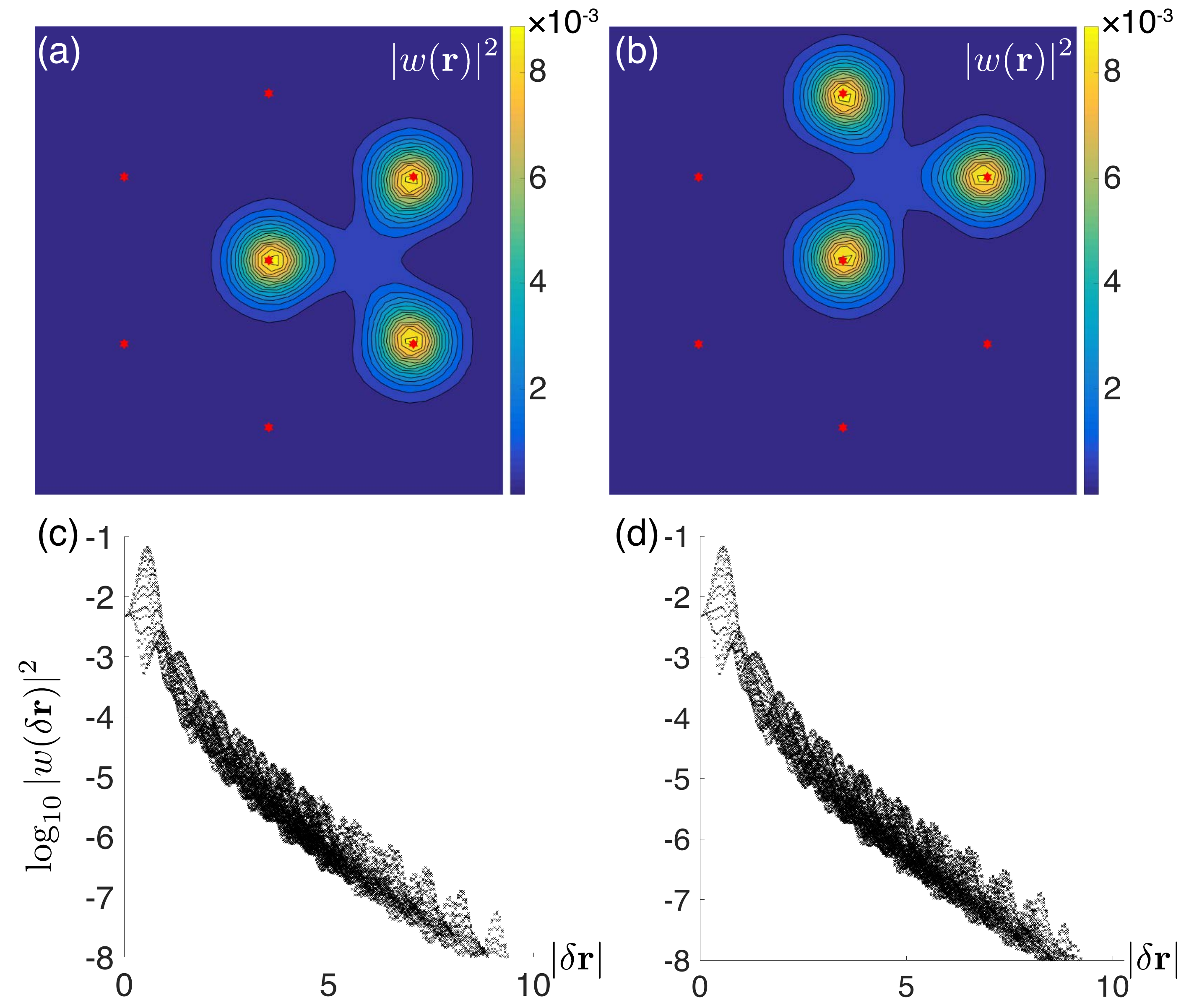}}
\caption{
{\bf Constructed Wannier functions.} As both valleys are taken into account, there are four nearly flat bands (spin ignored), giving rise to four Wannier functions per unit cell. Shown is for a pair related by $C_6$ symmetry; results for the other pair, which look nearly identical, are presented in Appendix \ref{app:Wannier}. (a,b) Amplitude of the Wannier function $w(\vec r)$. Red stars indicate the AA regions in the \moire potential. (c,d) The amplitude decays exponentially when the distance from the Wannier center, $\delta \vec r$ (measured in units of \moire lattice constant), is $\delta \vec r \gtrsim 3$.
\label{fig:WF}
 }
\end{center}
\end{figure}

Forgetting about valley conservation for the time-being, the construction of Wannier functions becomes a rather standard problem and well-established protocols apply. In particular, we construct well-localized Wannier functions using the projection method \cite{VanderbiltRMP}, starting from a set of well-localized trial wave-functions as the ``seed'' of the Wannier functions (Appendix \ref{app:Wannier}). 
Specifically, we will start with the continuum theory of Ref.\ \onlinecite{Bistritzer2011}, as described around Eq.\ \eqref{eq:CTh}, with the parameters $w_0 = 110$ meV and $w_1 = 120$ meV. The success of the construction hinges crucially on having a nonsingular projection everywhere in the BZ, which can be monitored by ensuring that the overlap between the seed and the actual Bloch wave-functions neither vanish nor diverge anywhere in the BZ \cite{VanderbiltRMP}.
Using this approach, we construct Wannier functions for a particular choice of parameters, detailed in Appendix \ref{app:Wannier}, for the four nearly flat bands near charge neutrality (spin ignored).
Our trial wave-functions attain a minimum and maximum overlap of $0.38$ and $3.80$ respectively, indicating a satisfactory construction.
Indeed, the Wannier functions we obtained are quite well-localized (Figs.\ \ref{fig:WF}(a,b)), with approximately $90 \%$ of their weight contained within one lattice constant from the Wannier center. In addition, the Wannier functions we constructed are exponentially localized (Figs.\ \ref{fig:WF}(c,d)), as anticipated from the nonsingular trial wave-function projection.

Having constructed the Wannier functions, one can readily extract an effective tight-binding model $\hat H^{\rm WF}$ by first projecting the full Bloch Hamiltonian into the Wannier basis, and then performing Fourier transform. Due to the exponential tail, however, the resulting tight-binding model would have infinite-range hopping despite with an exponentially suppressed amplitude. To capture the salient behavior of the model, it is typically sufficient to only keep the bonds with strength larger than some cutoff $t_c$. In other words, $t_c$ serves as a control parameter, and one recovers the exact band structure in the limit of $t_c \rightarrow 0$, albeit at the cost of admitting infinite-range hoppings.

The obtained band structures for different value of $t_c$ is plotted in Figs.\ \ref{fig:WFBands}(a--d). We found that a fairly long-range model (see Appendix \ref{app:TBModel} for parameters), keeping terms that connect sites up to $2$ lattice constants apart, is needed to capture all the salient features of the energetics. It should be noted that the range of the approximate models generally depends on the localization of the Wannier function, and in this work we have not optimized the Wannier functions. It is therefore possible that, by further optimization, one may capture the energetics more faithfully using only shorter-range terms.

Although spatial and time-reversal symmetries are respected in the tight-binding model, valley conservation is explicitly broken. This is because our Wannier functions cannot be chosen to represent valley conservation naturally, similar to the case of topological insulators \cite{Z2Wannier}, and therefore any truncation of the transformed Hamiltonian will generically introduce explicit valley-conservation symmetry breaking. Furthermore, one can ask how the operator $I_z$ is represented. In particular, we would want to construct the projection operators for the single-particle problem, ${\mathcal P}_{\pm} = \frac12(1\pm I_z)$, which project into the valleys. It would be most desirable if one can formulate $I_z$, and hence ${\mathcal P}_{\pm}$, directly in real-space. Given $I_z$ is also a free-electron Hermitian operator, this amounts to finding a symmetric Wannier representation of $I_z$. However, we found that there are again obstructions, which mirror exactly the obstructions we faced when attempting to construct Wannier functions for the single-valley two-band model of the nearly flat bands, i.e., a mismatch in the mirror eigenvalues, as well as a non-zero net charge of the Dirac points. Such inheritance of the obstructions is presumably a manifestation of the underlying anomaly of the single-valley description.

\begin{figure*}
{\includegraphics[width=0.95 \textwidth]{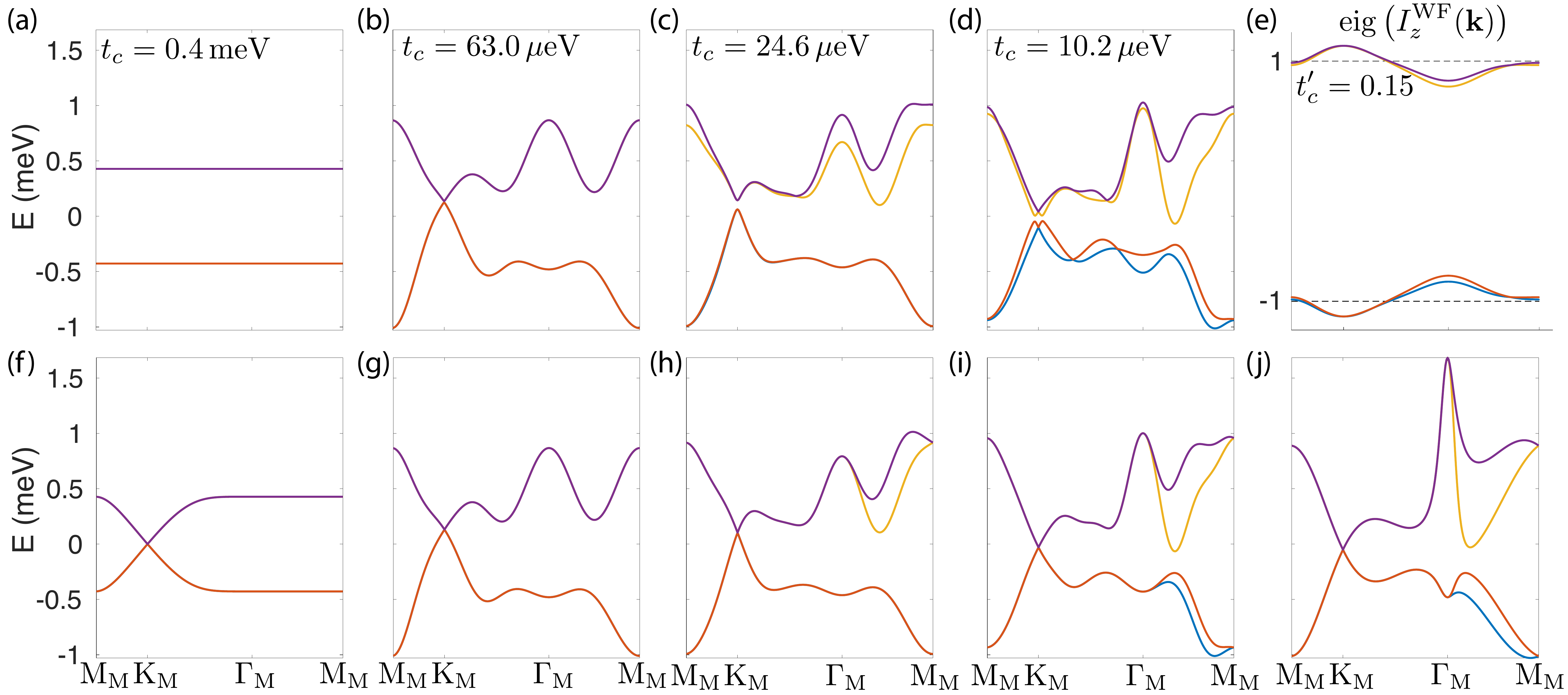}}
\caption{{Effective tight-binding model for the nearly flat bands.} (a-d) Effective tight-binding models for the nearly flat bands are derived by projecting the Hamiltonian in the continuum theory into the Wannier basis.
Bonds with strength $<t_c$ are truncated from the model, and the resulting band structures for three choices of $t_c$ are shown. (e) One can also derive the effective valley charge operator, $\hat I_z^{\rm WF}$, using the same procedure with cutoff $t_c'$. Similar to the Hamiltonian, truncation of terms in the effective operator will lead to error, and so the eigenvalues of $ I_z^{\rm WF} (\vec k)$ will have small deviation from the exact values of $\pm 1$. (f--i) Valley conservation can be re-enforced on the effective model through projection. The plots (f--i) correspond respectively to the effective models shown in (a--d). As $t_c$ is reduced, the effective model more faithfully reproduces the salient energetics features of the continuum theory, the latter of which is plotted in (j).
\label{fig:WFBands}
 }
\end{figure*}

{\em Towards recovering valley conservation:} It is desirable to restore valley conservation even approximately, for our truncated model, and we will describe a method below. Recall we denote the Hermitian valley charge operator in the continuum effective theory by $\hat I_z$ (Eq.\ \eqref{eq:IzDef}). Projecting $\hat I_z$ into the four-band subspace of the nearly flat bands, and rotating into the Wannier function basis, we arrive at another Hermitian operator defined on the honeycomb sites, which we can simply interpret as yet another Hamiltonian-like object in our problem. Similar to the earlier discussion for the Hamiltonian, the effective $\hat I_z$ operator, $\hat I_z^{\rm WF}$, will have infinite-range hopping with exponentially suppressed amplitude, and it is natural to approximate it by truncation, keeping again only terms with a strength larger than some $t_c'$. Such truncation, however, introduces deviation in the eigenvalues of $\hat I_z$ from the physical values of $\pm 1$ (Fig.\ \ref{fig:WFBands}e). To fix this, therefore, we can further perform spectral flattening of the corresponding bands to $\pm 1$ in momentum space. This procedure is well defined so long as a band gap is sustained between the second and third bands of $\hat I_z^{\rm WF}$, which is generically true as long as $t_c'$ is chosen to be reasonably small.
We will denote this flattened version by $\hat {\tilde I}_z^{\rm WF}$. Physically, this corresponds to an approximation of the actual representation of the valley charge operator in our tight-binding model, and again our approximation can be made exact in the limit of $t_c' \rightarrow 0$.

We can now restore valley conservation in our effective Hamiltonian. To this end, for $n\in \mathbb Z$ define the projection operator
\begin{equation}\begin{split}\label{eq:}
\hat {\mathcal P}_{n} = \sum_{\alpha} | n, \alpha \rangle \langle n, \alpha | ,
\end{split}\end{equation}
which projects into the sector with $\hat {\tilde I}_z^{\rm WF}$ eigenvalue $n$ (in the many-body Hilbert space) and satisfies $\hat {\mathcal P}_{n} \hat {\mathcal P}_{m} = \hat {\mathcal P}_{n} \delta _{n,m}$.
We can now define
\begin{equation}\begin{split}\label{eq:}
\hat {\tilde H}^{\rm WF} \equiv  \sum_{n \in \mathbb Z} \hat {\mathcal P}_n \hat H^{\rm WF} \hat {\mathcal P}_n,
\end{split}\end{equation}
for which valley conservation is restored. In essence, through this procedure we have introduced a pair of Hermitian operators $(\hat {\tilde H}^{\rm WF} , \hat {\tilde I}_z^{\rm WF})$, corresponding respectively to the Hamiltonian and the valley charge, that converge to the exact operators in the limit $t_c, t_c' \rightarrow 0$.

The valley projection procedure we described applies equally well to an interacting Hamiltonian $\hat H_{U}^{\rm WF}$ obtained by projecting the microscopic interaction terms to our Wannier basis, which again would not be automatically valley-conserving due to the truncation errors. For the free-part of the Hamiltonian, however, the projection procedure can be greatly simplified. This is because the Bloch states of $\hat {\tilde I}_z^{\rm WF}$, which equal to those of $\hat I_z^{\rm WF}$  by definition, are known, and using which we can decompose $H^{\rm WF}(\vec k)$ into the valley-conserving and valley-breaking parts. The projection then proceeds simply by retaining only the valley-conserving part. More concretely, write the Bloch ``Hamiltonian'' of the valley charge operator as $\tilde I_z(\vec k) = \Psi_{+; \vec k} \Psi^\dagger_{+; \vec k} - \Psi_{-; \vec k} \Psi^\dagger_{-; \vec k} $, where $\Psi_{\pm; \vec k}$ are $4\times 2$ matrices. Note that the columns of $\Psi_{\pm; \vec k}$ are simply the $\pm$-valley-charge eigenstates. We can then perform the projection by
\begin{equation}\begin{split}\label{eq:}
\tilde H^{\rm WF}(\vec k) =   \Psi_{+; \vec k}^\dagger  H^{\rm WF}(\vec k)  \Psi_{+; \vec k}
+  \Psi_{-; \vec k}^\dagger  H^{\rm WF}(\vec k)  \Psi_{-; \vec k},
\end{split}\end{equation}
giving an easy way to perform the described projection.

As is shown in Figs.\ \ref{fig:WFBands}(f--i), the projected effective tight-binding model re-exhibits all the symmetry features of the bands from the continuum theory (Fig.\ \ref{fig:WFBands}j), for any choice of truncation parameter $t_c$.
In particular, Figs.\ \ref{fig:WFBands}(a,f) represent the simplest model which demonstrates the utility of our approach, with the valley projection alone converting an otherwise hopping-free Hamiltonian into one exhibiting the charge-neutrality Dirac points.
We further remark that, although generically $H^{\rm WF}$ does not respect ${\rm U_v}(1)$, the effective Hamiltonian corresponding to Fig.\ \ref{fig:WFBands}b comes very close to being ${\rm U_v}(1)$-symmetric in terms of its the energetics along the high-symmetry line. We provide an explicit tabulation of the bonds in this $H^{\rm WF}$ and that of $I^{\rm WF}_z$ in Appendix \ref{app:TBModel}.

\subsection{Nearly flat bands in trilayer graphene-Boron Nitride \moire superlattices}
Recently, Mott insulating phases (but not superconductivity, at the time of writing) were observed \cite{Wang2018} in a heterostructure of ABC trilayer graphene encapsulated in boron nitride (TLG/hBN), where a \moire superlattice is present even at zero twisting between the graphene layers. Four mini bands are observed close to neutrality, whose bandwidth and separation can be tuned by a vertical electric field. Half-filling one of the nearly flat bands results in a Mott insulator.

We remark that the symmetry setup for this trilayer heterostructure bears more resemblance to the Bernal-stacked bilayer graphene than the TBG system we discussed above. In particular, the absence of Dirac points among the mini bands suggest that no $C_2$ symmetry is present, and the system is potentially described by a wallpaper group for which the two sublattices of the honeycomb lattice are no-longer symmetry related (say the wallpaper group $p3m1$, No.14 \cite{ITC}). If that is indeed the case, one expects the valley-resolved band structure to admit a tight-binding model defined on the triangular lattice, although it remains to be checked whether or not the charge density profile exhibits any nontrivial features, akin to that found for the TBG system (Fig.\ \ref{fig:WF}c). It will be of great interest to derive a concrete real-space effective model for the TLG/hBN, but we will leave this as a future work.

\section{Model for correlated states in trilayer graphene heterostructure} \label{sec: trilayer}

In this section we briefly consider the case of triangular \moire superlattices in trilayer graphene heterostructure. Correlated insulating states were observed very recently in this system \cite{Wang2018}.  Just like in the twisted bilayer here too there are nearly flat bands that are separated from the rest of the spectrum. However, unlike the TBG, here there are no Dirac crossings in this nearly flat band, and the low energy degrees of freedom are in the trivial representation of $C_3$. In addition, it is known that a vertical electric field can induce a gap for ABC-stacked trilayer graphene, and, depending on the direction of the electric field, the band structure can have a zero Chern for one direction of the electric field or a nonzero Chern number for the other direction of the electric field.\cite{Zhang2018} It is thus reasonable that, in the case where the direction of the electric field is such that the nearly flat bands posess no net Chern number, the nearly flat band can be modeled in real space by a triangular lattice model with two orbitals (corresponding to the two valleys) per site  supplemented with interactions.  However some care is still necessary. Time reversal and $C_2$ both act by flipping the valley index.  Thus the band dispersion $\epsilon_{\pm}(k)$  within a  single valley is not  symmetric under $k \rightarrow -k$:
\be
\epsilon_{\pm}(k) \neq \epsilon_{\pm}(-k)
\ee
though
\be
\epsilon_{+}(k) = \epsilon_{-}(-k)
\ee
is satisfied. A real space tight-binding description on the triangular lattice therefore takes the form
\be
H^t_{trilayer} = -\sum_{RR'} \sum_{a = \pm} \sum_{\alpha} t_{RR'} c^\dagger_{Ra\alpha} \left( e^{i\phi_{rr'} \tau_z} \right)_{aa'} c_{R'a'\alpha} + h.c
\ee
with $t_{rr'}$ real and positive. The phases $\phi_{rr'}$ are in general non-zero.  The phase $\phi$ even on nearest neighbor bonds cannot be removed as in general the symmetries permit a non-zero flux $\Phi$ for any single valley through an elementary triangle (and the opposite flux for the other valley).

When the Coulomb interaction dominates, the $SU(4)$ ferromagnet with a further selection of IVC order is once again a possibility. From a real space point of view the projection of the Coulomb interaction to the Wannier basis used to formulate the tight-binding model will lead  to an appropriate interaction Hamiltonian. If the Wannier orbitals are not tightly confined to each triangular site then there will be significant {\em inter-site} ferromagnetic Hund's exchange which will promote $SU(4)$ ferromagnetism with a further selection of IVC ordering.

It is interesting to consider the limit where the Wannier functions are sufficiently tightly localized that such a ferromagnetic inter-site exchange is weak and can be ignored. In that limit
to obtain a {\em minimal} model for this system we  restrict the hopping to just be nearest neighbor and include an on-site repulsion. The minimal model then takes the form
\begin{equation}\begin{split}\label{eq:}
H_{trilayer}  =&  H_0 + V \\
H_0  =&   -\sum_{\langle RR' \rangle} \sum_{a = \pm} \sum_{\alpha} t_{RR'} c^\dagger_{Ra\alpha} \left( e^{i\phi_{rr'} \tau_z} \right)_{aa'} c_{R'a'\alpha}\\
& + h.c \\
V  =&  \frac{U}{2}\sum_R(N_R-N_0)^2
\end{split}\end{equation}
with $\phi_{12} + \phi_{23} + \phi_{31} = \pm \Phi$ with $+$ sign for up-facing triangles and $-$ sign for downfacing ones. Here the sites $1, 2, 3$ are assumed to be arranged counterclockwise on each triangle. $N_R$ is the total electron number at site $R$, and $N_0$ controls the filling factor.
As in previous sections this Hamiltonian has a $U(2) \times U(2)$ symmetry corresponding to independent $U(2)$ rotations of each valley in addition to the discrete symmetries described above. The model thus needs to be supplemented with further weaker interactions that break the continuous symmetry down to $U(2)$ though we will not specify them here.  Note that if the flux $\Phi = 0$ then the model actually has an even higher $SU(4)$ symmetry.  Also, this model has an extra $C_2$ rotational symmetry that flips the valleys. This should be viewed as an emergent symmetry of the model defined above, which should be broken by other terms. In particular, it should be differentiated from a microscopic C2 symmetry; if such a microscopic symmetry was present, it would combine with time-rerveral symmetry and protect an odd number of Dirac points in the single-valley trilayer graphene band structure, which suffers from a parity anomaly.\cite{Koshino2009, Nandkishore2013} Our effective model does not suffer from this parity anomaly because this microscopic $C_2$ symmetry is absent.

The minimal model above allows for discussion of the Mott insulator in the strongly correlated regime of large $U$ at integer $N_0$. In the experiments Mott insulators at fillings $N_0 = 1,2$ have been reported. In the large-$U$ limit, the effective model takes the form of a
 ``spin-orbital"  Hamiltonian on a triangular lattice that has 4 states per site: 2 spins and 2  valleys.  A systematic $t/U$ expansion is readily performed to yield this spin-orbital Hamiltonian.
 At $O(\frac{t^2}{U})$ the ``super-exchange" is not sensitive to the flux $\Phi$, and we end up with an $SU(4)$ quantum antiferromagnet on the triangular lattice. For $N_0 = 1$ the $SU(4)$ spins are in the fundamental representation while for $N_0 = 2$ they are in the $6$-dimensional representation.

 Antiferromagnetic models of $SU(4)$ spins have been studied on a variety of lattices with different motivations (for some representative recent papers see Refs. \onlinecite{Corboz2012,Kaul2015,Yamada2017}). It seems likely that they go into ``paramagnetic" states that preserve $SU(4)$ symmetry.  However, a new feature in the present problem is the presence of the flux $\Phi$ in the underlying Hubbard model which breaks $SU(4)$ to $U(2) \times U(2)$.
 This modifies the spin-orbital model at $O(\frac{t^3}{U^2})$.  In the experiments the ratio of Coulomb interactions to the bandwidth of the nearly flat bands may be controlled by a perpendicular electric field, and it may be thus be possible to tune the strength of these third order terms relative to the second order ones. In Appendix \ref{app:SOMott} we derive the spin-orbital Hamiltonian to third order showing how the flux $\Phi$ leads to new terms not present in an $SU(4)$ invariant model. We however leave for the future a detailed study of these interesting spin-orbital models.

At any rate  we emphasize that this trilayer system is thus {\em qualitatively} different from the twisted bilayer graphene where we argued that a real space triangular lattice description is not possible due to Dirac crossings within the  nearly flat bands.

\section{Proposed future experiments} \label{sec: experimental proposal}

As discussed in previous sections, the ideas presented in this paper suggest a number of experiments which will be extremely useful in revealing the physics.  Here we reiterate and elaborate on some of these suggestions.

A crucial clue from the existing experiments is that an in-plane field suppresses the superconductivity - at optimal doping - when the Zeeman energy is of order the zero-field $T_c$. This indicates spin singlet pairing and that $T_c$ at optimal doping is associated with the loss of pairing. It will be extremely useful to study this systematically as a function of doping.   For the doped $C_3$ broken insulator, the superconductivity may be driven by pairing of a small Fermi surface of electrons. Then (except perhaps at very small doping) $T_c$ and the critical Zeeman scale will continue to track each other as the doping is decreased. In contrast, if the pairing (in the form of singlet valence bond formation)  already happens in the Mott insulator - as in the usual RVB theory,  or with the featureless Mott insulator, then with decreasing doping   $T_c$  and the critical Zeeman field should part ways significantly.

A second crucial clue from the experiments is the $2,4, 6,8,....$ degeneracy pattern of the Landau fan emanating from the Mott insulator.  We proposed that this was due to the freezing of the valley degree of freedom. This can be distinguished from the alternate possibility that there is spin freezing by studying the quantum oscillations in a tilted field. Zeeman splitting, if it exists, should  show up in a characteristic way as a function of tilt angle.

Our proposal is the intervalley phase coherence at a scale higher than both the superconducting $T_c \approx 1.5 K $ and the Mott insulating scale $\approx 5 K$.   The valley symmetry is as usual related to translational symmetry of the microscopic graphene lattice.  In the twisted bilayer there is an approximate translation symmetry that holds at some short scale associated with translation by one unit cell of the microscopic graphene lattices. Under this approximate translation operation electrons at the different valleys get different  phases. This is a $U_v(1)$ rotation.
Therefore intervalley ordering will strongly break this approximate short translation symmetry. Within each \moire site  the density of states will be uniform at the lattice scale when there is no intervalley ordering but will oscillate once this order sets in. This may be detectable through Scanning Tunneling Microscopy  (though if the bilayer graphene is fully encapsulated by Boron Nitride it may be challenging to  see the graphene layer).

Assuming there is intervalley ordering, if the undoped Mott insulator develops antiferromagnetic order, it appears likely that the doped superconductor will be a spin singlet $d_{x^2 - y^2} + i d_{xy}$ superconductor. This spontaneously breaks time reversal symmetry.  In contrast for a doped $C_3$ broken state, either $s$-wave or $d+id$ spin-singlet superconductivity seem possible.
It will also be useful to directly search for broken $C_3$ or \moire translational symmetry in the experiments.
Finally the very different behavior in quantum oscillations between electron and hole doping away from the Mott insulator suggests that there may be a first order transition into the Mott state as it is approached from the charge neutrality point. This will lead to hysteretic response as the gate voltage is tuned towards charge neutrality from the Mott insulator.

\section{Conclusion}
In this paper we addressed some of the theoretical challenges posed by the remarkable observations of Mott insulating states and proximate superconductivity in twisted bilayer graphene.

We proposed that both the Mott insulator and the superconductor develop out of a state with spontaneous intervalley coherence that breaks independent conservation of electrons at the two valleys.  We described  a mechanism for the selection of this order over other spin/valley polarized states owing to the peculiarities of the symmetry realization in the band structure. We showed that intervalley ordering by itself does not lead to a Mott insulator, and described possible routes through  which a Mott insulator can develop at low temperature. A specific concrete example is a $C_3$ broken insulator.  We showed how doping such an insulator  leads to an understanding of the quantum oscillation data, and presented a possible pairing mechanism for the development of superconductivity.  We described potentially useful experiments to distinguish the various possible routes to a Mott insulator from an intervalley coherent state.

Our work was  rooted in a microscopic understanding of the twisted graphene bilayer. We showed that   the momentum space structure of the nearly flat bands places strong constraints on real space descriptions. In particular, contrary to natural expectations,  we showed that a real space lattice model is necessarily  different from a correlated  triangular lattice model with two orbitals (corresponding to the two valleys) per site.  This is due to a symmetry enforced obstruction to constructing Wannier functions centered at the triangular sites that can capture the Dirac crossings of the nearly flat bands.  We showed that a honeycomb lattice representation may be possible but requires a non-local implementation of valley $U_v(1)$ symmetry.  In our description of the intervalley ordered state and its subsequent low temperature evolution into the Mott/superconducting states, we sidestepped these difficulties by first treating the problem directly in momentum space and defining a real space model only at scales below the intervalley ordering (when the obstruction to a honeycomb representation is gone).  We also contrasted the bilayer system with  trilayer graphene where Mott insulators have recently been observed. In the trilayer system, it is reasonable to construct a real space triangular lattice two-orbital model but the symmetries allow for complex hopping (with some restrictions). We argue that  this system may offer a valuable platform to realize interesting quantum spin-orbital liquids.

{\it Note added:}
After completion of this work, Ref. \onlinecite{Xu2018} appeared, which has significant differences from the present paper; after posting, Refs.\ \onlinecite{NoahLiang, Oskar, KoshinoLiang}, which discuss in particular the symmetries and constructions of Wannier functions, also appeared. Note that these discussions on Wannier functions disregard the presence of the (emergent) six-fold rotation symmetry, and, consequentially, the Dirac points observed at charge neutrality are not symmetry-protected features of these models.

\begin{acknowledgements}
We thank Yuan Cao, Valla Fatemi, and Pablo Jarillo-Herrero for extensive discussions of their data, and sharing their insights.  We also thank Shiang Fang, Liang Fu, Gene Mele, Patrick Lee, Oskar Vafek, Feng Wang, Noah F. Q. Yuan and Ya-Hui Zhang for interesting discussions or correspondence. TS is supported by a US Department of Energy grant
DE-SC0008739, and in part by a Simons Investigator award from the Simons Foundation. AV was supported by a Simons Investigator award and by NSF-DMR 1411343.
\end{acknowledgements}

\bibliography{TwBLGMott}

\clearpage
\onecolumngrid
\appendix

\section{Lattice and symmetries \label{app:GeoDetails}}
In this appendix, we document some details on the conventions and the symmetry transformations.

Consider a monolayer of graphene. We let the primitive lattice vectors $\vec A$ and reciprocal lattice vectors $\vec B$ be
\begin{equation}\begin{split}\label{eq:}
\vec A_1 = a \hat x, ~~~  \vec A_2 = a\left( - \frac{1}{2} \hat x + \frac{\sqrt{3}}{2} \hat y\right);~~~~~
\vec B_1 = \frac{4\pi}{ \sqrt{3} a} \left( \frac{\sqrt{3}}{2} \hat x + \frac{1}{2} \hat y \right), ~~~  \vec B_2 = \frac{4\pi}{ \sqrt{3} a} \hat y,
\end{split}\end{equation}
where $a = 2.46 \AA$ is the lattice constant (some authors use $a$ to denote the C-C bond length, which is a factor of $\sqrt {3}$ smaller than the lattice constant we are using here).
In this choice, we can choose the basis of the  honeycomb lattice sites to be
\begin{equation}\begin{split}\label{eq:}
\vec r_1 = \frac{1}{3} \vec A_1 + \frac{2}{3} \vec A_2; ~~~
\vec r_2 = \frac{2}{3} \vec A_1 + \frac{1}{3} \vec A_2.
\end{split}\end{equation}
In momentum space, the K, K' points are given by $ \pm  (\vec B_1 + \vec B_2)/3$, or, for the equivalent ones lying on the $x$-axis, $\pm (2 \vec B_1 - \vec B_2)/3$.
Note that $| \vec K| = 4\pi /(3a)$, as is well known. Furthermore, we take the Dirac speed $v_F$ to be $10^6 $ ms$^{-1}$.
Besides, we choose the \moire lattice vectors to be
\begin{equation}\begin{split}\label{eq:}
\vec a_1 = \frac{a}{2 \sin (\theta/2)} \left( \frac{\sqrt{3}}{2}, \frac{1}{2} \right); ~
\vec a_2 = \frac{a}{2 \sin (\theta/2)} \left( - \frac{\sqrt{3}}{2}, \frac{1}{2} \right).
\end{split}\end{equation}

In the main text, we have listed all the symmetries of the continuum theory (Table \ref{tab:Sym}). Here, we tabulate explicitly the symmetry transformations of the electron operators, which follow from that of the Dirac points in the monolayer problem.
\begin{equation}\begin{split}\label{eq:SymRep}
\hat t_{\vec \rho} \hat \psi_{\pm \mu;  \vec k} \hat t_{\vec \rho}^{-1} \propto \,&
e^{i  \vec k  \cdot \vec \rho} \hat \psi_{\pm  \mu;  \vec k};\\
\hat C_6 \hat \psi_{\pm  \mu;  \vec k} \hat C_6^{-1} =\,& \sigma_1 e^{\mp i \frac{2 \pi }{3}\sigma_3}\hat \psi_{\mp  \mu;  C_6 \vec k};\\
\hat M_y \hat \psi_{\pm  \mu;  \vec k} \hat M_y^{-1} =\,& \sigma_1  \hat \psi_{\pm  M_y[\mu];  M_y \vec k};\\
 \hat {\mathcal T} \hat \psi_{\pm  \mu;  \vec k} \hat {\mathcal T}^{-1} =\,&  \hat \psi_{\mp \mu; -  \vec k},
 \end{split}\end{equation}
where $\mu = {\rm t}, {\rm b}$. Note that $M_y $ is the only symmetry which flips the two layers, i.e., $M_y [{\rm t}] = {\rm b}$ and vice versa.

The symmetries listed in Eq.\ \eqref{eq:SymRep} generate all the spatial symmetries of the continuum theory of the TBG \cite{Neto2007,Bistritzer2011, Mele2011} (in wallpaper group 17). In particular, we see that $ t_{\vec \rho}$ and $M_y$ preserves the valley index (K vs.\ -K) whereas $C_6 $ and $\mathcal T$ do not.
However, their (pair-wise) nontrivial products will leave valley invariant, and it is helpful to also document their symmetry action explicitly (which are fixed by the above):
\begin{equation}\begin{split}\label{eq:SymRep2}
\hat C_3 \hat \psi_{\pm \mu;  \vec k} \hat C_3^{-1} =\,&  e^{\mp i \frac{2 \pi }{3}\sigma_3 }\hat \psi_{\pm \mu ;  C_3 \vec k};\\
(\hat C_6 \hat {\mathcal T}) \hat \psi_{\pm \mu;  \vec k} (\hat C_6 \hat {\mathcal T})^{-1} =\,& \sigma_1 e^{\pm i \frac{2 \pi }{3}\sigma_3}
\hat \psi_{\pm \mu; - C_6  \vec k};\\
(\hat C_2 \hat {\mathcal T}) \hat \psi_{\pm\mu;  \vec k} (\hat C_2 \hat {\mathcal T})^{-1} =\,& \sigma_1  \hat \psi_{\pm \mu;   \vec k}.
 \end{split}\end{equation}

Here, we make two remarks regarding the subtleties in the symmetry representation documented here: first, the momentum $\vec k$ appearing above are defined as the deviation from the original Dirac points in the monolayer problem. Generally, they correspond to different momenta in the \moire BZ. For instance, the Dirac point labeled by $(+, {\rm t})$, i.e., that of the $\vec K$ point in the top layer, is mapped to ${\rm K_M}$, whereas $(-, {\rm t})$ is mapped to ${\rm K_M'}$. Similarly, $(+, {\rm b})$ and $(-, {\rm b})$ are respectively mapped to ${\rm K_M'}$ and ${\rm K_M}$. If desired, one can also rewrite Eqs.\ \eqref{eq:SymRep} and \eqref{eq:SymRep2} using a common set of momentum coordinates defined with respect to the origin of the \moire BZ.

Second, the representation of the  translation symmetry, $\hat t_{\vec \rho}$, has a subtlety in its definition. This is because the microscopic translation effectively becomes an internal symmetry for the slowly varying fields appearing in the continuum theory. As such, for a single layer one can only deduce its representation up to an undetermined phase, and hence the appearance of $\propto $ in Eq.\ \eqref{eq:SymRep}. However, the relative momentum across the different slowly varying fields, say the operators corresponding to the $+$ valley of the top and bottom layers, is a physical quantity. Consequentially, there is really only one common ambiguity across all the degrees of freedom appearing in the continuum theory.

\section{Valley Symmetry and Wannier Obstruction \label{app:classAIII}}
We argue here that the valley symmetry resolved band structure does not admit a Wannier representation. Note, since we will ignore spin, this is a two-band model which will be crucial for what follows. If one were to include other bands the arguments below would fail, although precisely what selection of bands would lead to localized Wannier functions (LWFs) remains to be determined. In some ways, it is not very surprising that a valley resolved band structure does not admit a Wannier description, and a simple example is a single valley of monolayer graphene, which is just an isolated Dirac node. But in those cases the band structure does not terminate on raising the energy, and hence does not form an band. In contrast, in our present problem for TBG there is an isolated band, and so one may expect to capture the physics with LWFs. Nonetheless, we will argue there is an obstruction, as can be seen as follows.

We begin with three ingredients: (i) a two band model; (ii) $C_2{\mathcal T}$ symmetry; and a third ingredient which will be specified shortly. The two ingredients above enforce the following form on the momentum space Hamiltonian:
\begin{equation}
H = \epsilon_0(p)+ \epsilon_1(p) {\sigma_1} + \epsilon_2(p) {\sigma_2},
\end{equation}
where there is no condition on the function $\epsilon_i(p)$.
Similar to the main text, we implement $C_2{\mathcal T}$ by $\sigma_1 \mathcal K$, where $\mathcal K$ denotes complex conjugation. Check that $C_2{\mathcal T}$ leaves the Hamiltonian above  invariant. Now, if we are interested in the band wave-functions, they are independent of the first term in the hamiltonian, and we could pass to the following one by imposing a constraint:
\begin{equation}
H '=  \epsilon_1(p) {\sigma_1} + \epsilon_2(p) {\sigma_2}
\end{equation}
The obvious constraint is to demand:
\begin{equation}
\sigma_3 H'\sigma_3 = -H'.
\label{eq:chiral}
\end{equation}
This is nothing but the chiral condition that specifies class AIII. Now, we introduce the third ingredient: (iii) the two Dirac points at the middle of this band structure have the same chirality. This allows us to write down the following effective Hamiltonian close to neutrality:
\begin{equation}
H '_{low} =  -iv_F [\partial_x {\sigma_1}  + \partial_y {\sigma_2}]\otimes {\openone}
\end{equation}
Where we now have a four component structure to include the two Dirac nodes. Note, there is no mass term that will gap out these nodes and also preserve the chiral condition \ref{eq:chiral}, hence this corresponds to the surface of a three dimensional topological phase in class AIII, with index $\nu=2$, corresponding to the two Dirac nodes. Since this is the surface state of a nontrivial 3D topological phase, it does not admit a Wannier representation. However, when combined with the opposite valley band structure, together the pair of band structures does admit LWFs, but at the price of losing valley conservation symmetry.

Finally let us address a conundrum that the careful reader may be puzzled by. The valley resolved bands are stated to be the anomalous surface states of a 3D topological phase, nevertheless they appear as isolated bands which seems to contradict the usual expectation that such anomalous bands cannot be separated in energy. The way this is resolved in the present case is through the two band condition, which further allows us to map the problem to one with particle-hole symmetry (class AIII). The later problem can have anomalous surface states that are disconnected from the bulk bands because they are forced to stick at zero chemical potential, and hence cannot be deformed into the bulk. This mapping to class AIII only holds for the two band model, hence if we  add bands or fold the Brillouin zone from translation symmetry breaking, the presented arguments no longer hold.

\section{Wannier functions \label{app:Wannier}}

We will construct Wannier functions using the projection method \cite{VanderbiltRMP}. The method proceeds by first specifying a collection of well-localized, symmetric trial wave-functions in real space, which serves as the seed for constructing a smooth gauge required in obtaining well-localized Wannier functions for the problem of interest.

Let us begin by considering the symmetry properties of a real-space wave-function in our present problem.
For simplicity, we let $\beta$ be a collective index for the valleys and layers, i.e., $\beta = (+,{\rm t}), (-,{\rm t}), (+,{\rm b}), (-,{\rm b})$. Define the real-space electron operators
\begin{equation}\begin{split}\label{eq:psiR}
\hat \psi_{\beta; \vec r}^\dagger \propto & \int_0^{\Lambda} d^2 {\bar {\vec k}} \, e^{ - i {\bar {\vec k}} \cdot \vec r}  \hat \psi_{\beta;{\bar {\vec k}}}^\dagger,
\end{split}\end{equation}
where we will not keep track of the overall normalization of the operator. Here, ${\bar {\vec k}}$ is defined as the momentum measured with respect to the origin of the \moire BZ.
Note that $\hat \psi_{\beta; \vec r}^\dagger $ inherits symmetry transformation from that of  $\hat \psi_{\beta;{\bar {\vec k}}}^\dagger$.

To construct a collection of well-localized, symmetric trial wave-functions, one can follow the standard discussion concerning the symmetry representation associated with such real-space basis, say as reviewed in the supplemental materials of Ref.\ \onlinecite{NC}.
We will briefly sketch the main ideas below.
Let $ W^{\beta}_{\vec h_0}(\vec r)$ be a two-component (column) vector localized to $\vec h_0$ (the two-components here originate from the sublattice degree of freedom in the microscopic problem).
Define
\begin{equation}\begin{split}\label{eq:WExpand}
\hat W_{\vec h_0}^{ \dagger} \equiv
\sum_\beta  \int d^2{\vec r}\,\hat \psi_{\beta;\vec r}^\dagger  W^{\beta}_{\vec h_0}(\vec r),
\end{split}\end{equation}
and its associated momentum-space operator
\begin{equation}\begin{split}\label{eq:}
 \hat \Gamma^{ \dagger }_{\vec h_0 ; \vec k}  \equiv  \frac{1}{V} \sum_{\vec a} e^{i \vec k \cdot  \vec a }  \hat W_{\vec h_0  + \vec a}^{\dagger }
= \int d^2 \vec k \, \hat \psi_{\beta; \vec k} \, \Gamma^{\beta }_{\vec h_0} ( \vec k),
\end{split}\end{equation}
where $\vec a$ is a \moire lattice vector, and we assume a periodic system with $V$ \moire unit cells.

For our purpose, we want $\hat W_{\vec h_0}^{ \dagger} $ to serve as our seed for the construction of symmetric Wannier functions. To this end, suppose $\vec h_0 = (\vec a_1 - \vec a_2)/3$, which corresponds to a honeycomb site in the \moire potential, i.e., an AB/BA region. We demand $\hat W_{\vec h_0}^{ \dagger} $ to be invariant under time-reversal, the mirror $M_y$, and the three-fold rotation about $\vec h_0$, which we denote by $\bar C_3$.
In addition, recall that the previously predicted charge density profile \cite{Magaud2010, ShiangPRB,STMPRL, STMNatPhy,Crommie2015} suggests that the Wannier functions will take the shape shown in Fig.\ \ref{fig:WF}a. Therefore, it is natural to consider a trial  $\hat W_{\vec h_0}^{ \dagger} $ taking the form
\begin{equation}\begin{split}\label{eq:}
\hat W_{\vec h_0}^\dagger = \hat w_{\vec 0}^\dagger + \hat {\bar C}_3 \hat w_{\vec 0}^\dagger  \hat {\bar C}_3^\dagger + \hat {\bar C}_3^2  \hat w_{\vec 0}^\dagger  \hat {\bar C}_3^{2\dagger},
\end{split}\end{equation}
where $\hat w_{\vec 0}^\dagger$ is localized to the unit-cell origin (an AA site). By definition, $\hat W_{\vec h_0}^\dagger$ transforms trivially under $\bar C_3$, which one can verify would lead to the correct $C_3$ representations for the nearly flat bands.

It remains to ensure that $\hat w_{\vec 0}^\dagger $ is symmetric under $\mathcal T$ and $M_y$. In the spirit of Eq.\ \eqref{eq:WExpand}, we may write $\hat w_{\vec 0}^\dagger = \sum_\beta \hat w_{\vec 0}^{\beta \dagger}$ for $\beta = (+,{\rm t}), (-,{\rm t}), (+,{\rm b}), (-,{\rm b})$. From the symmetry transformation in Eq.\ \eqref{eq:SymRep}, we set
\begin{equation}\begin{split}\label{eq:}
\hat w_{\vec 0}^{-{\rm t} \dagger} = \hat {\mathcal T}   \hat w_{\vec 0}^{+{\rm t} \dagger}  \hat {\mathcal T} ^{-1};~~~
\hat w_{\vec 0}^{-{\rm b} \dagger} = \hat {\mathcal T}   \hat w_{\vec 0}^{+{\rm b} \dagger}  \hat {\mathcal T} ^{-1}.
\end{split}\end{equation}
Similarly, to respect $M_y$ symmetry we set
\begin{equation}\begin{split}\label{eq:}
\hat w_{\vec 0}^{+{\rm b} \dagger} = \zeta_{M_y}\,  \hat M_y \hat w_{\vec 0}^{+{\rm t} \dagger}  \hat M_y^{-1},
\end{split}\end{equation}
where $\zeta_{M_y}=\pm 1$. As such, we have reduced our degree of freedom on the specification of the trial wave-function  $\hat W_{\vec h_0}^\dagger$ down to our choice of $\hat w_{\vec 0}^{+{\rm t} \dagger} =  \int d^2 \vec r\,  \hat \psi_{\beta, \vec r}^\dagger\,w^{+{\rm t} }_{\vec 0}(\vec r) $. Our only condition on $w^{+{\rm t} }_{\vec 0}(\vec r)$ is that it is a two-component wave-function well-localized to $\vec 0$, however, we will simply assume a Gaussian form:
\begin{equation}\begin{split}\label{eq:WFGauss}
w^{+ {\rm t}}_{\vec 0}(\vec r) =& e^{- | \vec r |^2/2 \xi^2} \phi_{\vec 0}^{+ {\rm t}},
\end{split}\end{equation}
where $\xi$ is the localization length, and $ \phi_{\vec 0}$ is a constant two-component vector.
Correspondingly, all the other two-component wave-function will take a similar form, although they can be localized to a different point, say $\bar C_3 \vec 0$.
Such a choice is particularly convenient, as its Fourier transform can be readily evaluated:
\begin{equation}\begin{split}\label{eq:SimpleGam}
\gamma^\beta_{\vec c}(\vec k) =  \int d^2 \vec r \, e^{- i \vec k \cdot \vec r} e^{- |\vec r - \vec c |^2/2 \xi^2} \phi_{\vec c}^\beta
\propto  \, e^{- i \vec k\cdot \vec c} e^{-|\vec k|^2 \xi^2/2} \phi_{\vec c}^\beta,
\end{split}\end{equation}
which enables an efficient computation of the overlap between our trial  and the Bloch wave-functions.

Thus far, we have focused on only one well-localized wave-function in real space. In our problem, the two sites of the effective honeycomb lattice are related by $C_6$, i.e., we simply construct the wave-function localized to $\vec h_1 \equiv C_6 \vec h_0$ through
\begin{equation}\begin{split}\label{eq:SimpleGam}
\hat W_{\vec h_1}^\dagger \equiv \hat C_6 \hat W_{\vec h_0}^\dagger \hat C_6^{-1}.
\end{split}\end{equation}
Besides, to describe a four-band problem we should have two orbitals on each of the honeycomb sites. These two orbitals are not symmetry-related. However, to reproduce the $M_y$ representation in momentum space, we have to take the two orbitals to respectively correspond to $\zeta_{M_y}=+1$ and $\zeta_{M_y}=-1$.

\begin{figure}
\begin{center}
{\includegraphics[width=0.48 \textwidth]{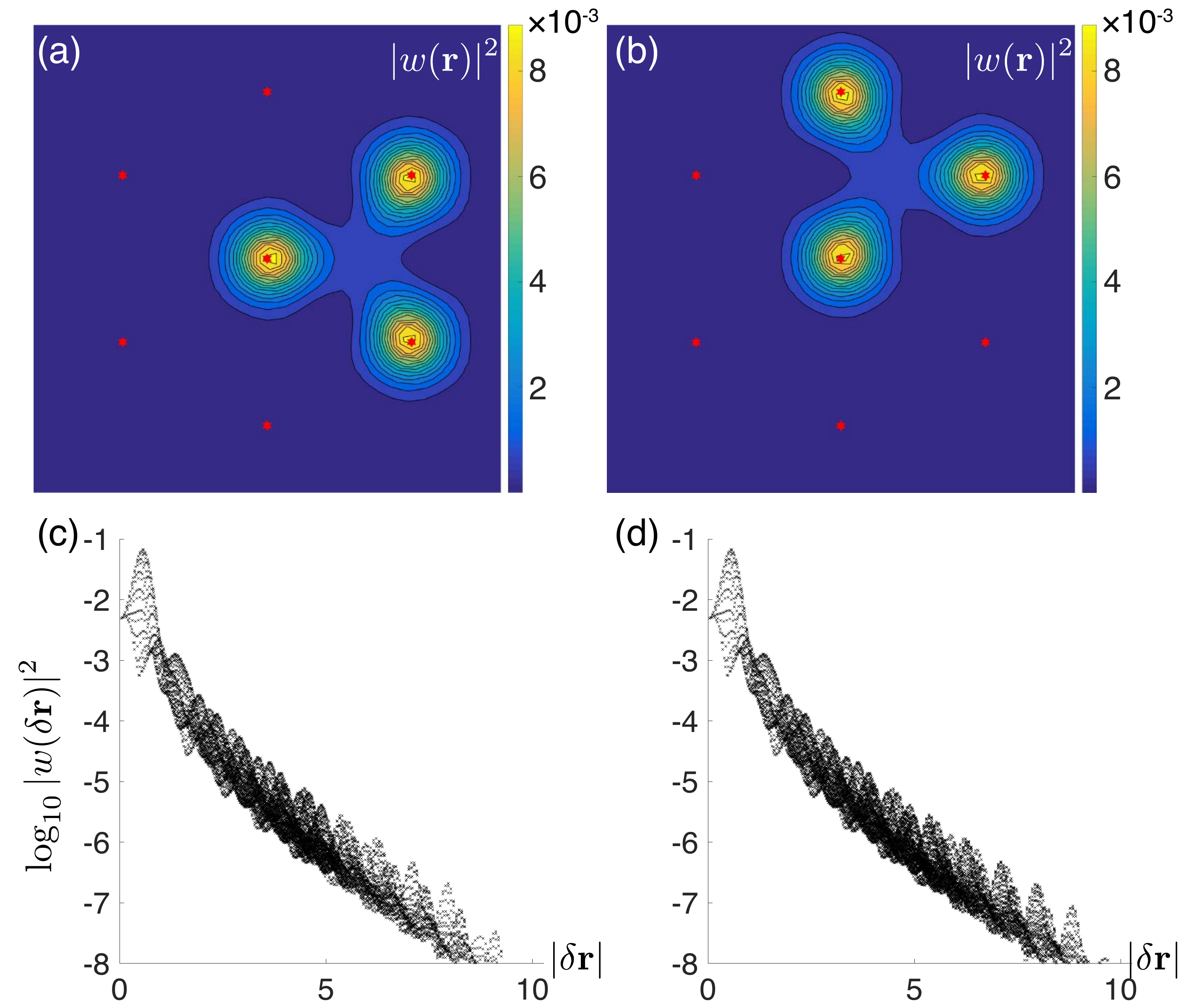}}
\caption{{\bf Localization of the other set of constructed Wannier functions.}
\label{fig:WFSupp}
 }
\end{center}
\end{figure}

In our numerical construction of the Wannier functions, we take the localization length to be $0.15 |\vec a_1|$, and the two-component vectors
\begin{equation}\begin{split}\label{eq:}
\phi^{+{\rm t},\zeta_{M_y}=+1}_{\vec 0} = \left(
\begin{array}{c}
-0.416 + 0.168 i\\
0.820 + 0.356i
\end{array}
\right);~~~
\phi^{+{\rm t},\zeta_{M_y}=-1}_{\vec 0} = \left(
\begin{array}{c}
0.296  -0.380 i\\
0.776  + 0.407i
\end{array}
\right).
\end{split}\end{equation}
These choices are found simply by a search of the parameter space to optimize the minimum overlap between the trial and the Bloch wave-functions.
We check the overlap for $\sim 1180$ momenta along the high-symmetry line ${\rm M_M}$--${\rm K_M}$--$\Gamma_{\rm M}$--${\rm M_M}$, as well as for an additional $1000$ randomly sampled points in the BZ. The minimum and maximum over the overlap are respectively found to be $0.38$ and $3.80$, indicating a satisfactory construction of the Wannier functions. In Fig.\ \ref{fig:WF} of the main text, we present the results for the two symmetry-related Wannier functions labeled by $\zeta_{M_y}=+1$; the corresponding results for the pair with $\zeta_{M_y}=-1$ is shown in FIg.\ \ref{fig:WFSupp}. Remarkably, the localization property of the two sets are essentially identical.

\section{Tight-binding model \label{app:TBModel}}
In this appendix, we provide an explicit tabulation of the bonds in the $H^{\rm WF}$ corresponding to Figs.\ \ref{fig:WFBands}(b,g) in the main text, as well as the associated $I^{\rm WF}_z$ ,whose eigenvalues are plotted in Fig.\ \ref{fig:WFBands}e.

In the following, we parametrize a ``bond'' by
\begin{equation}\begin{split}\label{eq:}
t \, \hat c_{\vec r_{\rm To} + \vec a}^\dagger \hat c_{\vec r_{\rm Fr}} + {\rm h.c.},
\end{split}\end{equation}
where $\vec a$ is a \moire lattice vector, and ${\rm To}$, ${\rm Fr} = 1,\dots ,4$ labels the four Wannier functions localized to each unit cell. Physically, orbital $1$ corresponds to the one localized to $\vec h_0$ with $\zeta_{M_y}=+1$; orbital $2$ is the one  localized to $\vec h_1$ symmetry-related to orbital $1$; orbital $3$ is localized to $\vec h_0$ with $\zeta_{M_y}=-1$; and orbital $4$ is symmetry-related to $3$.

The bonds in $H^{\rm WF}$  with $t_c = 63\,\mu{\rm eV}$ are tabulated in Table \ref{tab:EffHIz}a. Note that we only tabulated half of the bonds, in the sense that the Hermitian conjugate of the listed bonds are not included. In particular, for consistency we have also halved the coefficient of the on-site chemical potential $\sim t \hat c^\dagger_{\vec r} \hat c_{\vec r}$, which is Hermitian by itself. We have in addition subjected the ``trace part'' of the chemical potential (i.e., we have removed a constant energy offset) from the model.
The effective valley charge operator defined with $t_c' = 0.15$ is similarly tabulated in Table  \ref{tab:EffHIz}b.

\begin{center}
\begin{table}
\caption{{\bf Effective Hamiltonian $H^{\rm WF}$ and valley charge operator $I^{\rm WF}_z$}. We denote a lattice vector $\vec a = l_1 \vec a_1 + l_2 \vec a_2$ by $(l_1,l_2)$.
\label{tab:EffHIz}}
\begin{tabular}{cccr}
~\\
\multicolumn{4}{c}{(a) $H^{\rm WF}$}\\
${\rm To}$ & ${\rm Fr}$ & $\vec a$ & $t\, (\mu{\rm eV})$\\
\hline \hline
$1$ & $1$ & $(0,0)$  & $213.8$\\
$2$ & $2$ & $(0,0)$  & $213.8$\\
$3$ & $3$ & $(0,0)$  & $-213.8$\\
$4$ & $4$ & $(0,0)$  & $-213.8$\\
$2$ & $3$ & $(0,1)$  & $-76.2$\\
$4$ & $1$ & $(0,1)$  & $-76.2$\\
$1$ & $1$ & $(0,2)$  & $81.2$\\
$1$ & $4$ & $(0,2)$  & $76.2$\\
$2$ & $2$ & $(0,2)$  & $81.2$\\
$3$ & $2$ & $(0,2)$  & $76.2$\\
$3$ & $3$ & $(0,2)$  & $-67.2$\\
$4$ & $4$ & $(0,2)$  & $-67.2$\\
$2$ & $3$ & $(1,-1)$  & $-76.2$\\
$3$ & $3$ & $(1,-1)$  & $66.4$\\
$4$ & $1$ & $(1,-1)$  & $-76.2$\\
$4$ & $4$ & $(1,-1)$  & $66.4$\\
$2$ & $3$ & $(1,1)$  & $76.2$\\
$4$ & $1$ & $(1,1)$  & $76.2$\\
$3$ & $3$ & $(1,2)$  & $66.4$\\
$4$ & $4$ & $(1,2)$  & $66.4$\\
$1$ & $1$ & $(2,0)$  & $81.2$\\
$2$ & $2$ & $(2,0)$  & $81.2$\\
$3$ & $3$ & $(2,0)$  & $-67.2$\\
$4$ & $4$ & $(2,0)$  & $-67.2$\\
$1$ & $4$ & $(2,1)$  & $76.2$\\
$3$ & $2$ & $(2,1)$  & $76.2$\\
$3$ & $3$ & $(2,1)$  & $66.4$\\
$4$ & $4$ & $(2,1)$  & $66.4$\\
$1$ & $1$ & $(2,2)$  & $81.2$\\
$1$ & $4$ & $(2,2)$  & $-76.2$\\
$2$ & $2$ & $(2,2)$  & $81.2$\\
$3$ & $2$ & $(2,2)$  & $-76.2$\\
$3$ & $3$ & $(2,2)$  & $-67.2$\\
$4$ & $4$ & $(2,2)$  & $-67.2$\\
\hline \hline
\end{tabular}
~~~~~~~~~~~~~~~~~~~~~~~~~~~~~~~~~~~~~~~~
\begin{tabular}{cccr}
~\\
\multicolumn{4}{c}{(b) $I_z^{\rm WF}$}\\
${\rm To}$ & ${\rm Fr}$ & $\vec a$ & $t$~~~~~~~\\
\hline \hline
$1$ & $2$ & $(0,0)$  & $0.451\,i$\\
$3$ & $4$ & $(0,0)$  & $-0.443\,i$\\
$1$ & $2$ & $(0,1)$  & $0.451\,i$\\
$1$ & $3$ & $(0,1)$  & $-0.217\,i$\\
$2$ & $4$ & $(0,1)$  & $-0.217\,i$\\
$3$ & $1$ & $(0,1)$  & $-0.217\,i$\\
$3$ & $4$ & $(0,1)$  & $-0.443\,i$\\
$4$ & $2$ & $(0,1)$  & $-0.217\,i$\\
$1$ & $2$ & $(1,0)$  & $-0.172\,i$\\
$1$ & $3$ & $(1,0)$  & $-0.217\,i$\\
$2$ & $1$ & $(1,0)$  & $0.172\,i$\\
$2$ & $4$ & $(1,0)$  & $-0.217\,i$\\
$3$ & $1$ & $(1,0)$  & $-0.217\,i$\\
$3$ & $4$ & $(1,0)$  & $0.181\,i$\\
$4$ & $2$ & $(1,0)$  & $-0.217\,i$\\
$4$ & $3$ & $(1,0)$  & $-0.181\,i$\\
$1$ & $2$ & $(1,1)$  & $0.451\,i$\\
$1$ & $3$ & $(1,1)$  & $0.217\,i$\\
$2$ & $4$ & $(1,1)$  & $0.217\,i$\\
$3$ & $1$ & $(1,1)$  & $0.217\,i$\\
$3$ & $4$ & $(1,1)$  & $-0.443\,i$\\
$4$ & $2$ & $(1,1)$  & $0.217\,i$\\
$1$ & $2$ & $(1,2)$  & $-0.172\,i$\\
$3$ & $4$ & $(1,2)$  & $0.181\,i$\\
\hline \hline\\
~\\~\\~\\~\\~\\~\\~\\~\\~\\
\end{tabular}
\end{table}
\end{center}

\clearpage

\section{Hartree-Fock theory for selection of IVC ordering \label{app:HF}}
Here we discuss a simple mean field treatment to illustrate that an IVC state is favored by the system at $\nu=-2$, which is described by the following simplified Hamiltonian of Eqn. (\ref{HFHam}):
\beq \label{eq: HF-hamiltonian}
H=H_0+V
\eeq
where the free Hamiltonian is
\beq \label{eq: HF-kinetic}
H_0=\sum_{an\alpha\vec k}\epsilon_{an}(\vec k)c_{an\alpha}^\dag(\vec k) c_{an\alpha}(\vec k)
\eeq
with $a$ the valley index, $n$ the band index, and $\alpha$ the spin index. Notice that the dispersion $\epsilon_{an}(\vec k)$ is independent of the spin, and due to time reversal $\epsilon_{an}(\vec k)=\epsilon_{-an}(-\vec k)$.

We assume a simple form of interaction:
\beq \label{eq: HF-interaction-simplified}
V=\frac{g}{N}\sum_{\vec k_1\vec k_2\vec q}c^\dag_{an\alpha}(\vec k_1+\vec q)c_{an\alpha}(\vec k_1)c^\dag_{a'n'\alpha'}(\vec k_2-\vec q)c_{a'n'\alpha'}(\vec k_2)
\eeq
where $n_{a\alpha}(x)$ is the electron density with flavor $a$ and spin $\alpha$. Repeated indices are summed over here. This interaction has an $SU(8)$ symmetry. As discussed in the maintext, the more complete form of the  interaction that takes into account the form factors arising from projecting the interactions onto the nearly flat bands should be 
\beq \label{eq: HF-interaction-projected}
\begin{split}
	V&=\frac{g}{N}\sum_{\vec k_1\vec k_2\vec q}\Lambda^{a}_{nn'}(\vec k_1+\vec q, \vec k_1)\Lambda^{a'}_{mm'}(\vec k_2-\vec q, \vec k_2)\\
	&\cdot c^\dag_{an\alpha}(\vec k_1+\vec q)c_{an'\alpha}(\vec k_1)
	\cdot c^\dag_{a'm\alpha'}(\vec k_2-\vec q)c_{a'm'\alpha'}(\vec k_2)
\end{split}
\eeq
with the form factors given by the Bloch wave functions of the states in the nearly flat bands via
\beq
\Lambda^a_{nn'}(\vec k_1, \vec k_2)=\la u_{an}(\vec k_1)|u_{an'}(\vec k_2)\ra
\eeq
where $|u_{an}(\vec k)\ra$ is the Bloch wave function of a state in the nearly flat bands labelled by valley index $a$, band index $n$ and momentum $\vec k$ (it has no dependence on the spin indices). However, for simplicity, we will first present the result from analysing the simplified interaction (\ref{eq: HF-interaction-simplified}), and comment on the preliminary result from analysing (\ref{eq: HF-interaction-projected}) at the end of this appendix.

We will factorize the interaction in a Hartree-Fock mean field manner
\beq
\begin{split}
&V_{\rm MF}\\
=&\frac{g}{N}\sum_{\vec k_1\vec k_2\vec q}
\Big[
\la c^\dag_{an\alpha}(\vec k_1+\vec q)c_{an\alpha}(\vec k_1)\ra
c^\dag_{a'n'\alpha'}(\vec k_2-\vec q)c_{a'n'\alpha'}(\vec k_2)
+\la c^\dag_{a'n'\alpha'}(\vec k_2-\vec q)c_{a'n'\alpha'}(\vec k_2)\ra c^\dag_{an\alpha}(\vec k_1+\vec q)c_{an\alpha}(\vec k_1)\\
&-\la c^\dag_{an\alpha}(\vec k_1+\vec q)c_{a'n'\alpha'}(\vec k_2)\ra c^\dag_{a'n'\alpha'}(\vec k_2-\vec q)c_{an\alpha}(\vec k_1)
-\la c^\dag_{a'n'\alpha'}(\vec k_2-\vec q)c_{an\alpha}(\vec k_1)\ra c^\dag_{an\alpha}(\vec k_1+\vec q)c_{a'n'\alpha'}(\vec k_2)\\
&-\la c^\dag_{an\alpha}(\vec k_1+\vec q)c_{an\alpha}(\vec k_1)\ra
\la c^\dag_{a'n'\alpha'}(\vec k_2-\vec q)c_{a'n'\alpha'}(\vec k_2)\ra
+\la c^\dag_{an\alpha}(\vec k_1+\vec q)c_{a'n'\alpha'}(\vec k_2)\ra\la c^\dag_{a'n'\alpha'}(\vec k_2-\vec q)c_{an\alpha}(\vec k_1)\ra
\Big]
\end{split}
\eeq

The first, second and fifth terms are the Hartree contributions, while the other terms are the Fock contributions. The Hartree contribution is determined by the local total electron density alone and independent of ordering, so for our purposes they can be simply dropped. We will thus focus on the Fock terms.

We would like to compare the energies of a spin polarized state (SP), a valley-Z-polarized state (IzP) and a valley-XY-polarized state (IVC). In state SP, we assume
\beq
\la c^\dag_{an\alpha}(\vec k_1)c_{a'n\alpha'}(\vec k_2)\ra=\delta_{aa'}\delta_{nn'}(n_{1n}(\vec k_1)\delta_{\alpha\alpha'} +\phi_{1n}(\vec k_1)\sigma^z_{\alpha\alpha'})\delta_{\vec k_1\vec k_2}
\eeq
The corresponding macroscopic quantities are defined as
\beq
n_{1n}=\frac{1}{N}\sum_{\vec k}n_{1n}(\vec k),
\quad
\phi_{1n}=\frac{1}{N}\sum_{\vec k}\phi_{1n}(\vec k)
\eeq
In state IzP, we assume
\beq
\la c^\dag_{an\alpha}(\vec k_1)c_{a'n'\alpha'}(\vec k_2)\ra=\delta_{nn'}\delta_{\alpha\alpha'}(n_{2n}(\vec k_1)\delta_{aa'} +\phi_2(\vec k_1)\tau^z_{aa'})\delta_{\vec k_1\vec k_2}
\eeq
The corresponding macroscopic quantities are defined as
\beq
n_{2n}=\frac{1}{N}\sum_{\vec k}n_{2n}(\vec k),
\quad
\phi_2=\frac{1}{N}\sum_{\vec k}\phi_2(\vec k)
\eeq
In the IVC state, we assume
\beq
\la c^\dag_{an\alpha}(\vec k_1)c_{a'n'\alpha'}(\vec k_2)\ra=\delta_{nn'}\delta_{\alpha\alpha'} (n_{3an}(\vec k_1)\delta_{aa'}+\phi_{3n}(\vec k_1)\tau^x_{aa'})\delta_{\vec k_1\vec k_2}
\eeq
Notice here $\phi_{3n}$ is a complex number, and changing its phase implies changing the valley order in the valley-XY-plane. For simplicity, we will take $\phi_{3n}$ to be positive for both $n$. The corresponding macroscopic quantities are defined as
\beq
n_{3an}=\frac{1}{N}\sum_{\vec k}n_{3an}(\vec k),
\quad
\phi_{3n}=\frac{1}{N}\sum_{\vec k}\phi_{3n}(\vec k)
\eeq

Below we calculate the energies of these states by the mean field approximation.

\begin{enumerate}
\item
{\em Spin polarized state}

We start with the SP state. In this case, the interaction Hamiltonian is replaced by its mean field representative, which reads
\beq
V_{\rm MF}=g\sum_{\vec k}[-2\phi_{1n}\sigma^z_{\alpha\alpha'}c^\dag_{an\alpha}(\vec k)c_{an\alpha'}(\vec k)+4(\phi_{11}^2+\phi_{12}^2 -n_{11}^2-n_{12}^2)]
\eeq
The total mean field Hamiltonian is then
\beq
\begin{split}
&H_{\rm MF}\\
=&H_0+V_{\rm MF}\\
=&\sum_{an\vec k}\big((\epsilon_{an}(\vec k)-2g\phi_{1n})c^\dag_{an+}(\vec k) c_{an+}(\vec k)+(\epsilon_{an}(\vec k)+2g\phi_{1n})c^\dag_{an-}(\vec k) c_{an-}(\vec k)\big)+4gN(\phi_{11}^2+\phi_{12}^2-n_1^2-n_2^2)
\end{split}
\eeq

Consider the limit where $g$ is much larger than the bandwidth. In this limit we expect the spin is fully polarized. Without loss of generality, assume $\phi_{1n}\geqslant 0$ for both $n$. In this case, all electrons will be in the state with $\alpha=+$ and $n=1$. Self-consistency requires that
\beq
\begin{split}
\frac{1}{N}\sum_{\vec k}\la c^\dag_{a1+}(\vec k)c_{a'1+}(\vec k)\ra=\delta_{aa'}=\delta_{aa'}(n_{11}+\phi_{11})\\
\frac{1}{N}\sum_{\vec k}\la c^\dag_{a1-}(\vec k)c_{a'1-}(\vec k)\ra=0=\delta_{aa'}(n_{11}-\phi_{11})\\
\frac{1}{N}\sum_{\vec k}\la c^\dag_{a2+}(\vec k)c_{a'2+}(\vec k)\ra=0=\delta_{aa'}(n_{12}+\phi_{12})\\
\frac{1}{N}\sum_{\vec k}\la c^\dag_{a2-}(\vec k)c_{a'2-}(\vec k)\ra=0=\delta_{aa'}(n_{12}-\phi_{12})\\
\end{split}
\eeq
which yields
\beq
n_{11}=\phi_{11}=\frac{1}{2},
\quad
n_{12}=\phi_{12}=0
\eeq
This is indeed a fully polarized state.

In this fully polarized state, and further assuming that the lower flat band is strictly lower than the higher flat band, the total energy of this state is
\beq \label{eq: energy of state-1}
E_1=\sum_{a\vec k}\epsilon_{a1}(\vec k)-2gN
\eeq

\item
{\em Valley $I^z$ polarized}

Next we turn to the state IzP. In this case, the interaction Hamiltonian is replaced by
\beq
\begin{split}
V_{\rm MF}
&=g\sum_{\vec k}(-2\phi_2\tau^z_{aa'}c^\dag_{an\alpha}(\vec k)c_{a'n\alpha}(\vec k)+8\phi_2^2-4(n_{21}^2+n_{22}^2))
\end{split}
\eeq
The total mean field Hamiltonian is
\beq
\begin{split}
&H_{\rm MF}\\
=&H_0+V_{\rm MF}\\
=&\sum_{n\alpha\vec k}\big((\epsilon_{+n}(\vec k)-2g\phi_2)c^\dag_{+n\alpha}(\vec k) c_{+n\alpha}(\vec k)+(\epsilon_{-n}(\vec k)+2g\phi_2)c^\dag_{-n\alpha}(\vec k) c_{-n\alpha}(\vec k)\big)+gN(8\phi_2^2-4(n_{21}^2+n_{22}^2))
\end{split}
\eeq

Suppose $g$ is much larger than the band width, the system wants to fully polarize along $a=+$, and both bands with $n=1$ and $n=2$ will be occupied for $a=+$. Self-consistency requires that
\beq
\begin{split}
\frac{1}{N}\sum_{\vec k}\la c^\dag_{+1\alpha}(\vec k)c_{+1\alpha'}(\vec k)\ra=\delta_{\alpha\alpha'}=\delta_{aa'}(n_{21}+\phi_{2})\\
\frac{1}{N}\sum_{\vec k}\la c^\dag_{+2\alpha}(\vec k)c_{+2\alpha'}(\vec k)\ra=\delta_{\alpha\alpha'}=\delta_{aa'}(n_{22}+\phi_{2})\\
\frac{1}{N}\sum_{\vec k}\la c^\dag_{-1\alpha}(\vec k)c_{-1\alpha'}(\vec k)\ra=0=\delta_{aa'}(n_{21}-\phi_{2})\\
\frac{1}{N}\sum_{\vec k}\la c^\dag_{-2\alpha}(\vec k)c_{-2\alpha'}(\vec k)\ra=0=\delta_{aa'}(n_{22}-\phi_{2})\\
\end{split}
\eeq
which yields
\beq
n_{21}=n_{22}=\phi_2=\frac{1}{2}
\eeq
This is indeed a fully valley-Z-polarized state.

In this fully polarized case, the total energy of this state is
\beq \label{eq: energy of state-2}
E_2=2\sum_{\vec k}\epsilon_{+1}(\vec k)-2gN=E_1
\eeq

\item
{\em IVC state}

Finally we discuss the IVC state. In this case, the interaction Hamiltonian is replaced by
\beq \label{eq: mean field interaction for state-3}
\begin{split}
V_{\rm MF}
=g\sum_{\vec k}(-2\phi_{3n}(c^\dag_{+n\alpha}(\vec k)c_{-n\alpha}(\vec k)+c^\dag_{-n\alpha}(\vec k)c_{+n\alpha}(\vec k))+2(2\phi_{3n}^2 -n_{3an}^2))
\end{split}
\eeq
The total mean field Hamiltonian is
\beq
\begin{split}
&H_{\rm MF}\\
=&H_0+V_{\rm MF}\\
=&\sum_{an\alpha\vec k}\epsilon_{an}(\vec k)c_{an\alpha}^\dag(\vec k) c_{an\alpha}(\vec k)-2g\sum_{n\alpha\vec k}\phi_{3n}
\big(
c^\dag_{+n\alpha}(\vec k)c_{-n\alpha}(\vec k)
+c^\dag_{-n\alpha}(\vec k)c_{+n\alpha}(\vec k)
\big)+2gN(2\phi_{3n}^2-n_{3an}^2)
\end{split}
\eeq

Denote $\bar\epsilon_n(\vec k)=\frac{\epsilon_{+n}(\vec k)+\epsilon_{-n}(\vec k)}{2}$ and $\delta\epsilon_n(\vec k)=\frac{\epsilon_{+n}(\vec k)-\epsilon_{-n}(\vec k)}{2}$, the spectrum of the above Hamiltonian is
\beq
E_{\pm n}(\vec k)=\bar{\epsilon}_n(\vec k)\pm\sqrt{\delta\epsilon_n(\vec k)^2+4g^2\phi_{3n}^2}
\eeq
Denote the eigenmodes corresponding to $E_{\pm n}(\vec k)$ by $d_{\pm n}(\vec k)$, they satisfy
\beq
\left(
\begin{array}{c}
	c_{+n}(\vec k)\\
	c_{-n}(\vec k)
\end{array}
\right)
=
\left(
\begin{array}{cc}
	\cos\frac{\theta_n(\vec k)}{2} & -\sin\frac{\theta_n(\vec k)}{2} \\
	\sin\frac{\theta_n(\vec k)}{2} & \cos\frac{\theta_n(\vec k)}{2}
\end{array}
\right)
\left(
\begin{array}{c}
	d_{+n}(\vec k)\\
	d_{-n}(\vec k)
\end{array}
\right)
\eeq
with
\beq
\cos\theta_n(\vec k)=\frac{\delta\epsilon_n(\vec k)}{\sqrt{\delta\epsilon_n(\vec k)^2+4g^2\phi_{3n}^2}},
\quad
\sin\theta_n(\vec k)=\frac{-2g\phi_{3n}}{\sqrt{\delta\epsilon_n(\vec k)^2+4g^2\phi_{3n}^2}}
\eeq

Again consider the limit where $g$ is much larger than the bandwidth, then system tend to occupy the bands with energies $E_{-1}$. Self-consistency requires that
\beq
\begin{split}
\phi_{3n}&=\frac{1}{N}\sum_{\vec k}\la c^\dag_{+n\alpha}(\vec k)c_{-n\alpha}(\vec k)\ra
=\sum_{\vec k}\frac{\sin\theta_n(\vec k)}{2N}\la d^\dag_{+n\alpha}(\vec k)d_{+n\alpha}(\vec k)-d^\dag_{-n\alpha}(\vec k)d_{-n\alpha}(\vec k)\ra\\
&=\frac{1}{N}\sum_{\vec k}\frac{g\phi_{3n}}{\sqrt{\delta\epsilon_n(\vec k)^2+4g^2\phi_{3n}^2}}\la
d^\dag_{-n\alpha}(\vec k)d_{-n\alpha}(\vec k)-d^\dag_{+n\alpha}(\vec k)d_{+n\alpha}(\vec k)\ra\\
n_{3+n}&=\frac{1}{N}\sum_k\left(\frac{1+\cos\theta_n(\vec k)}{2}\la d^\dag_{+n\alpha}(k)d_{+n\alpha}(k)\ra+ \frac{1-\cos\theta_n(\vec k)}{2}\la d^\dag_{-n\alpha}(k)d_{-n\alpha}(k)\ra\right)\\
n_{3-n}&=\frac{1}{N}\sum_k\left(\frac{1+\cos\theta_n(\vec k)}{2}\la d^\dag_{-n\alpha}(k)d_{-n\alpha}(k)\ra+ \frac{1-\cos\theta_n(\vec k)}{2}\la d^\dag_{+n\alpha}(k)d_{+n\alpha}(k)\ra\right)
\end{split}
\eeq
In the limit where $g$ is much larger than the bandwidth, $\phi_{3n}\rightarrow 1/2$, which means the system tends to fully polarize along the valley-XY direction. At the same time, $n_{an}\rightarrow 1/2$. The total energy of this state in this case is
\beq
E_3=2\sum_{k}(\bar\epsilon_1(k)-\sqrt{\delta\epsilon_1(k)^2+g^2})
\eeq
This energy is lower than $E_2$ in (\ref{eq: energy of state-2}):
\beq
\begin{split}
E_3-E_2
&=2\sum_{k}(\bar\epsilon_1(k)-\sqrt{\delta\epsilon_1(k)^2+g^2}-\epsilon_{+1}(\vec k)+g)\\
&=2\sum_k(g-\delta\epsilon_1(k)-\sqrt{\delta\epsilon_1(k)^2+g^2})<0
\end{split}
\eeq
where time reversal symmetry of the non-interacting band structure has been used in the last step: $\sum_k\delta\epsilon_1(k)=0$.

 \end{enumerate}

Therefore, if the interaction strength is much larger than the bandwidth, the system tends to be fully polarized. Within this analysis, among different fully polarized states, the valley-XY-ordered state (IVC) has lower energy than a spin-polarized state and a valley-Z-ordered state.

Finally, we comment on the effect of taking into account the form factors into the interaction term, as in (\ref{eq: HF-interaction-projected}). It turns out that in this case which state has the lowest energy can potentially be changed from the IVC state if bands are too flat, and our premininary calculations show that the spin-polarized state and the valley-Z-ordered state will have a lower energy compared to the IVC state when the interaction strength is around 10 times of the bandwidth of the nearly flat bands. A more reliable way to settle down the real ground state is to do a numerical calculation formulated in momentum space, using the Hamiltonian given by (\ref{eq: HF-hamiltonian}), (\ref{eq: HF-kinetic}) and (\ref{eq: HF-interaction-projected}).

\section{Spin-orbital model for Mott insulators in trilayer graphene \label{app:SOMott}}

In this appendix we derive an effective spin-orbital model applicable to the trilayer graphene described in Sec. \ref{sec: trilayer}, in the limit of $U\gg t$. This effective model will be applicable for both the case of $\nu=-1$ and $\nu=-2$, and it is obtained by a systematic expansion in the large-$U$ limit to the order of $t^3/U^2$.

Besides the translation symmetry of the triangular lattice, the system is assumed to have $U(2)\times U(2)$ symmetries, corresponding to charge and flavor $U(1)$ conservations, as well as spin conservation. In addition, there is a $C_6$ symmetry that maps one flavor into the other, and a time reversal symmetry that also maps one flavor into the other (while leaving the spin unchanged, so $T^2=1$ for this time reversal).

We start from a Hubbard model with nearest-neighbor hopping and on-site Hubbard interactions. Consider three nearby sites (A, B, and C) forming an elementary triangle of this triangular lattice. The kinetic Hamiltonian is taken as
\beq
H_0=-t\sum_{a\alpha}
e^{i\eta_a\phi}
\left(c^\dag_{a\alpha}(B)c_{a\alpha}(A)
+c^\dag_{a\alpha}(C)c_{a\alpha}(B)
+c^\dag_{a\alpha}(A)c_{a\alpha}(C)
\right)+\hc
\eeq
where $a=\pm$ labels the flavor, $\alpha$ labels the spin, and $\eta_a=\pm 1$ if $a=\pm$. The interaction Hamiltonian on each site is taken as
\beq
V=\frac{U}{2}(N-N_0)^2
\eeq
where $N=\sum_{a\alpha}c^\dag_{a\alpha}(\vec r_i)c_{a\alpha}(\vec r_i)$, and we are interested in both the cases with $N_0=2$ and $N_0=1$. All other terms of the many-body system can be generated by applying various symmetries.

We are looking for an effective spin-orbital model at large $U$ to the order of $t^3/U^2$. We first present the result and discuss some simple physical consequences before presenting the details of the derivation. The final result is
\beq
H_{\rm eff}=H^{(2)}+H^{(3)}
\eeq
with
\beq
H^{(2)}=\frac{2t^2}{U}\frac{1}{n^2}(n^2+\vec S_A\cdot\vec S_B)(n^2+\vec I_A\cdot\vec I_B)-\frac{4nt^2}{U}
\eeq
and
\beq
H^{(3)}=\frac{3t^3}{U^2}\left[-\frac{1}{4n^3}h^{(3,1)}+\frac{1}{4n^2}h^{(3,2)}\right]-\frac{12nt^3}{U^2}
\eeq
where
\beq
\begin{split}
	h^{(3,1)}
	=&8n^2\cos 3\phi(n^2+\vec S_A\cdot\vec S_B+\vec S_B\cdot\vec S_C+\vec S_C\cdot\vec S_A)(n^2+I_A^zI_B^z+I_B^zI_C^z+I_C^zI_A^z)\\
	+&4n^2(n^2+\vec S_A\cdot\vec S_B+\vec S_B\cdot\vec S_C+\vec S_C\cdot\vec S_A)\cdot\left[e^{i\phi}(I_A^+I_B^-+I_B^+I_C^-+I_C^+I_B^-)+\hc\right]\\
	+&8\sin 3\phi[I_A^zI_B^zI_C^z+n^2(I_A^z+I_B^z+I_C^z)]\cdot\left[\vec S_A\cdot(\vec S_B\times\vec S_C)\right]\\
	-&4i\vec S_A\cdot(\vec S_B\times\vec S_C)\cdot
	\left[
	I_A^z(e^{i\phi}I_B^+I_C^--e^{-i\phi}I_B^-I_C^+)+(A\rightarrow B\rightarrow C\rightarrow A)
	\right]
\end{split}
\eeq
and
\beq
\begin{split}
	h^{(3,2)}
	&=T_B^{kj}T_C^{jk}e^{2i\phi(I_B^z+2I_C^z)}
	+e^{-2i\phi(I_B^z+2I_C^z)}T_B^{jk}T_C^{kj}
	+(B\rightarrow C\rightarrow A\rightarrow B)\\
	&=4\cdot\left(n^2+\vec S_A\cdot\vec S_B\right)\cdot\left(2n^2\cos 3\phi+2I_A^zI_B^z\cos 3\phi+e^{-i\phi}I_A^+I_B^-+e^{i\phi} I_A^-I_B^+\right)+(B\rightarrow C\rightarrow A\rightarrow B)
\end{split}
\eeq
In the above $n=\frac{1}{2}$ for $N_0=1$, and $n=1$ for $N_0=2$.

We briefly comment on these results before turning to the detailed derivation. First, we notice the effective model indeed has the same set of symmetries as the original Hubbard model. Second, we note $H^{(2)}$ is actually $SU(4)$ invariant, and $H^{(3)}$ is also $SU(4)$ invariant at $\phi=0$, which is most easily seen by inspecting  (\ref{eq: 3rd order N0=1}) and (\ref{eq: 3rd order N0=2}). These $SU(4)$ symmetric interactions can potentially make the system an $SU(4)$ antiferromagnet. Third, there is a spin chirality term for the two valleys with opposite coefficients, which can potentially lead to interesting kinds of topological order, such as a double-semion state. Lastly, we note the system may develop an $SU(4)$ antiferromagnetic order in the limit where $U\gg t$ at $N_0=2$, although there is evidence that the system is disordered at $N_0=1$. Suppose the $SU(4)$ antiferromagnetic Heisenberg model indeed results in an $SU(4)$-broken state, we would like to understand how the $SU(4)$-breaking terms in the Hamiltonian affects the ground state.

To this end, we expand $H^{(3)}$ for small $\phi$ and obtain
\beq
\begin{split}
h^{(3,1)}
=&8n^2(n^2+\vec S_A\cdot\vec S_B+\vec S_B\cdot\vec S_C+\vec S_C\cdot\vec S_A)(n^2+\vec I_A\cdot\vec I_B+\vec I_B\cdot\vec I_C+\vec I_C\cdot\vec I_A)\\
&-8[\vec S_A\cdot(\vec S_B\times\vec S_C)]\cdot[\vec I_A\cdot(\vec I_B\times\vec I_C)]\\
&+24\phi[I_A^zI_B^zI_C^z+n^2(I_A^z+I_B^z+I_C^z)]\cdot\left[\vec S_A\cdot(\vec S_B\times\vec S_C)\right]\\
&-8\phi n^2(n^2+\vec S_A\cdot\vec S_B+\vec S_B\cdot\vec S_C+\vec S_C\cdot\vec S_A)(I_A^yI_B^x-I_A^xI_B^y+(A\rightarrow B\rightarrow C\rightarrow A))\\
&+4\phi\vec S_A\cdot(\vec S_B\times\vec S_C)\cdot
\left[
I_A^z(I_B^+I_C^-+I_B^-I_C^+)+(A\rightarrow B\rightarrow C\rightarrow A)\right]+\mc{O}(\phi^2)
\end{split}
\eeq
and
\beq
\begin{split}
h^{(3,2)}
=&8(n^2+\vec S_A\cdot\vec S_B)(n^2+\vec I_A\cdot\vec I_B)+(A\rightarrow B\rightarrow C\rightarrow A)\\
&+8\phi(n^2+\vec S_A\cdot\vec S_B)(I_A^yI_B^x-I_A^xI_B^y)+(A\rightarrow B\rightarrow C\rightarrow A)+\mc{O}(\phi^2)
\end{split}
\eeq
As we can see, the main effect of $SU(4)$-breaking interactions to the leading order of $\phi$, besides giving rise to the spin chirality terms, is to introduce terms roughly in the following form to the Hamiltonian
\beq
\delta H=J\phi (I_A^yI_B^x-I_A^xI_B^y+(A\rightarrow B\rightarrow C\rightarrow A))
\eeq
Consider the $I$'s as classical spins on the XY-plane and parametrize $I^x_A=\cos\theta_A$ and $I_y^A=\sin\theta_A$, the above Hamiltonian becomes
\beq
\delta H=J\phi\sin(\theta_A-\theta_B)+(A\rightarrow B\rightarrow C\rightarrow A))
\eeq
This term tends to introduce some canting of the $I$ order to gain energy.

 Below we will derive effective spin-orbital Hamiltonians of such systems in the large-$U$ limit to the order of $t^3/U^2$, for $N_0=1$ and $N_0=2$ separately. We will use the standard Van-Vleck perturbation theory to derive the effective Hamiltonian, which involves two steps: writing down the matrix elements of the effective Hamiltonian within the low-energy manifold, and expressing these matrix elements in terms of charge neutral operators.

In this problem, the effective Hamiltonian can be written as
\beq
H_{\rm eff}=H^{(2)}+H^{(3)}+\cdots
\eeq
with
\beq
H^{(2)}=\mc{P}H_0\mc{D}H_0\mc{P}
\eeq
and
\beq
H^{(3)}=\mc{P}H_0\mc{D}H_0\mc{D}H_0\mc{P}
\eeq
where $\mc{P}$ is the projector into the ground state manifold of $V$, and
\beq
\mc{D}=\frac{1-\mc{P}}{E_0-V}
\eeq

\subsection{Mott insulator at $N_0=1$} \label{sec: N0=1}

We first discuss the effective Hamiltonian for $N_0=1$. We will first write the results in terms of some $SU(4)$ generators, and convert them into a form in terms of the $\vec S$ and $\vec I$ operators later. The final result in terms of $SU(4)$ generators is
\beq \label{eq: 2nd order N0=1}
H^{(2)}=\frac{2t^2}{U}\sum_{i,j=1}^4T_A^{ij}T_B^{ji}-\frac{2t^2}{U}
\eeq
where
\beq
T^{ij}=c_i^\dag c_j
\eeq
with $i=1, 2, 3, 4$, and $|1\ra=|+\uparrow\ra$, $|2\ra=|+\downarrow\ra$, $|3\ra=|-\uparrow\ra$ and $|4\ra=|-\downarrow\ra$. In terms of these states, the action of $T^{ij}$ is
\beq
T^{ij}=\delta_{jk}|i\ra
\eeq

And
\beq \label{eq: 3rd order N0=1}
\begin{split}
	H^{(3)}
	=&-\frac{6t^3}{U^2}\sum_{i,j,k=1}^{4}
	\left[
	T_A^{ji}T_B^{kj}T_C^{ik}e^{-2i\phi(I_A^z+I_B^z+I_C^z)}
	+e^{2i\phi(I_A^z+I_B^z+I_C^z)}T_A^{ij}T_B^{jk}T_C^{ki}
	\right]\\
	&+\frac{3t^3}{U^2}\sum_{j,k=1}^{4}
	\left[
	T_B^{kj}T_C^{jk}e^{2i\phi(I_B^z+2I_C^z)}
	+e^{-2i\phi(I_B^z+2I_C^z)}T_B^{jk}T_C^{kj}
	+(B\rightarrow C\rightarrow A\rightarrow B)
	\right]
	-\frac{6t^3}{U^2}\cos3\phi
\end{split}
\eeq

The details of the calculations are below.

\subsubsection{Order $t^2/U$}

To get the effective Hamiltonian at the order $t^2/U$, it is sufficient to consider a single bond of the triangular lattice, and all other terms can be generated by applying symmetries.

Now we first calculate the matrix elements of the effective Hamiltonian at the order $t^2/U$. Denote the two sites linked by this bond by $A$ and $B$, then $|i,j\ra=c^\dag_i(A)c^\dag_j(B)|0\ra$.

There are only two types of nonzero matrix elements:
\beq
\begin{split}
\la i, j|H^{(2)}|i, j\ra&=-\frac{2t^2}{U}\\
\la j, i|H^{(2)}|i, j\ra&=\frac{2t^2}{U}
\end{split}
\eeq
for $i\neq j$.

These matrix elements can be recast into the effective Hamiltonian
\beq
H^{(2)}=\frac{2t^2}{U}\sum_{i,j=1}^4T_A^{ij}T_B^{ji}-\frac{2t^2}{U}
\eeq
where
\beq
T^{ij}=c_i^\dag c_j
\eeq
such that
\beq
T^{ij}|k\ra=\delta_{jk}|i\ra
\eeq
These operators satisfy the commutation relation of $SU(4)$ generators:
\beq \label{eq: su(4) algebra}
[T^{ij}, T^{kl}]=\delta_{jk}T^{il}-\delta_{il}T^{kj}
\eeq

\subsubsection{Order $t^3/U^2$}

Now we turn to the order $t^3/U^2$. A simple inspection of the model shows that, to calculate the effective Hamiltonian at the order of $t^3/U^2$ in this problem, we only need to consider a single elementary triangle, then all other terms can be obtained by symmetries.

Now we first calculate the matrix elements of the effective Hamiltonian to the order $t^3/U^2$. To this end, we need to first have a systematic way to label the states in the ground state manifold of the three-site problem of a single elementary triangle. It turns out we can classify the states into 3 classes: $|i, j, k\ra$, $|i, i, k\ra$ and $|i, i, i\ra$ with $i\neq j\neq k\neq i$, such that the matrix elements between states from two different classes always vanish.  Now we only need to calculate the matrix elements between states within the same classes.

The results are

\beq
\la i, i, i|H^{(3)}|i, i, i\ra=0
\eeq

\beq
\begin{split}
	\la i,i,j|H^{(3)}|i,i,j\ra&=-\frac{t^3}{U^2}(-2\cos\eta_i\phi+2\cos\eta_j\phi)=0\\
	\la i,j,i|H^{(3)}|i,i,j\ra&=\frac{3t^3}{U^2}(e^{i\eta_i\phi+2i\eta_j\phi}-e^{-i\eta_j\phi-2i\eta_i\phi})
\end{split}
\eeq

\beq
\begin{split}
	\la i,j,k|H^{(3)}|i,j,k\ra&=-\frac{2t^3}{U^2}(\cos 3\eta_i\phi+\cos 3\eta_j\phi+\cos 3\eta_k\phi) =-\frac{6t^3}{U^2}\cos3\phi\\
	\la i,k,j|H^{(3)}|i,j,k\ra&=\frac{3t^3}{U^2}(e^{2i\eta_k\phi+i\eta_j\phi}+e^{-2i\eta_j\phi-i\eta_k\phi})\\
	\la j,k,i|H^{(3)}|i,j,k\ra&=-\frac{6t^3}{U^2}e^{-i\eta_i\phi-i\eta_j\phi-i\eta_k\phi}\\
	\la k,i,j|H^{(3)}|i,j,k\ra&=-\frac{6t^3}{U^2}e^{i\eta_i\phi+i\eta_j\phi+i\eta_k\phi}
\end{split}
\eeq

Recasting these matrix elements into a compact form yields
\beq
\begin{split}
	H^{(3)}
	=&-\frac{6t^3}{U^2}\sum_{i,j,k=1}^{4}
	\left[
	T_A^{ji}T_B^{kj}T_C^{ik}e^{-2i\phi(I_A^z+I_B^z+I_C^z)}
	+e^{2i\phi(I_A^z+I_B^z+I_C^z)}T_A^{ij}T_B^{jk}T_C^{ki}
	\right]\\
	&+\frac{3t^3}{U^2}\sum_{j,k=1}^{4}
	\left[
	T_B^{kj}T_C^{jk}e^{2i\phi(I_B^z+2I_C^z)}
	+e^{-2i\phi(I_B^z+2I_C^z)}T_B^{jk}T_C^{kj}
	+(B\rightarrow C\rightarrow A\rightarrow B)
	\right]
	-\frac{6t^3}{U^2}\cos3\phi
\end{split}
\eeq

\subsection{Mott insulator at $N_0=2$} \label{N0=2}

Now we discuss the effective Hamiltonian for $N_0=2$. In this case there are 6 states on each site, which can be denoted as $|ij\ra\equiv c^\dag_ic^\dag_j|0\ra=-|ji\ra$.

Again we will first write the result in terms of some $SU(4)$ generators, and then convert it to a form in terms of the original $\vec S$ and $\vec I$ operators. The final result is
\beq \label{eq: 2nd order N0=2}
H^{(2)}=\frac{2t^2}{U}\sum_{ij}T_A^{ij}T_B^{ji}-\frac{4t^2}{U}
\eeq
where $T^{ij}=c_i^\dag c_j$. And
\beq \label{eq: 3rd order N0=2}
\begin{split}
	H^{(3)}
	=&-\frac{6t^3}{U^2}
	\left[
	T_A^{lj}T_B^{jk}T_C^{kl}e^{2i\phi(\tilde{I}_A^z+\tilde{I}_B^z+\tilde{I}_C^z)}
	+e^{-2i\phi(\tilde{I}_A^z+\tilde{I}_B^z+\tilde{I}_C^z)}T_A^{jl}T_B^{kj}T_C^{lk}
	\right]\\
	&+\frac{3t^3}{U^2}
	\left[
	T_{B}^{ij}T_C^{ji}e^{2i\phi(\tilde{I}_B^z+2\tilde{I}_C^z)}+\hc
	+(B\rightarrow C\rightarrow A\rightarrow B)
	\right]
	-\frac{12t^3}{U^2}\cos3\phi
\end{split}
\eeq
where $\tilde{I}_A^z$ gives the flavor of the particle that is acted by the $T$ operators. For example, $T_A^{lj}T_B^{jk}T_C^{kl} e^{2i\phi(\tilde{I}_A^z+\tilde{I}_B^z+\tilde{I}_C^z)} =T_A^{lj}T_B^{jk}T_C^{kl}e^{i\phi(\eta_j+\eta_k+\eta_l)} =e^{2i\phi(\tilde{I}_A^z+\tilde{I}_B^z+\tilde{I}_C^z)}T_A^{lj}T_B^{jk}T_C^{kl}$.

The details of the calculations are below.

\subsubsection{Order $t^2/U$}

To get the effective Hamiltonian at the order $t^2/U$, it is sufficient to consider a single bond of the triangular lattice, and all other terms can be generated by applying symmetries.

Now we first calculate the matrix elements of the effective Hamiltonian at the order of $t^2/U$. Dnote the two sites linked by this bond by $A$ and $B$, then $|ij, kl\ra=c^\dag_i(A)c^\dag_j(A)c^\dag_k(B)c^\dag_l(B)|0\ra$.

It is useful to distinguish 3 types of states: $|ij, kl\ra$, $|ij, ik\ra$ and $|ij, ij\ra$, where different letters denote different states. Clearly there is no matrix elements between two states from two different types, and all we need is to calculate the matrix elements between states within the same type.

The independent matrix elements include

\beq
\la ij, ij|H^{(2)}|ij, ij\ra=0
\eeq

\beq
\begin{split}
\la ij, ik|H^{(2)}|ij, ik\ra&=-\frac{2t^2}{U}\\
\la ij, ik|H^{(2)}|ik, ij\ra&=\frac{2t^2}{U}
\end{split}
\eeq

\beq
\begin{split}
	\la ij, kl|H^{(2)}|ij, kl\ra&=-\frac{4t^2}{U}\\
	\la ik, jl|H^{(2)}|ij, kl\ra&=\frac{2t^2}{U}
\end{split}
\eeq
All other matrix elements at this order either vanish or can be obatined from the above by permutations.

From these matrix elements we obtain $H^{(2)}$:
\beq
H^{(2)}=\frac{2t^2}{U}\sum_{ij}T_A^{ij}T_B^{ji}-\frac{4t^2}{U}
\eeq
As in the case of $N_0=1$, $T^{ij}=c_i^\dag c_j$ that satisfies (\ref{eq: su(4) algebra}), the commutation relations of the generators of $SU(4)$. Now acting on 2-particle states on each site, the actions of these operators are
\beq
T^{ij}|kl\ra=\delta_{jk}|il\ra+\delta_{jl}|ki\ra
\eeq

\subsubsection{Order $t^3/U^2$}

Now we turn to the order $t^3/U^2$. A simple inspection of the model shows that, to calculate the effective Hamiltonian at the order of $t^3/U^2$ in this problem, we only need to consider a single elementary triangle, then all other terms can be obtained by symmetries.

Now we first calculate the matrix elements of the effective Hamiltonian to the order $t^3/U^2$. To this end, we need to first have a systematic way to label the states in the ground state manifold of the three-site problem of a single elementary triangle.

It turns out there are 5 types of states: $|ij, ij, ij\ra$, $|ij, jk, ij\ra$, $|ij, jk, ik\ra$, \{$|ij, kl, il\ra$, $|ij, kl, ij\ra$\} and $|ij, ik, il\ra$, where, for example,
\beq
|ij, jk, ki\ra=c^\dag_i(A) c^\dag_j(A) c^\dag_j(B) c^\dag_k(B) c^\dag_k(C) c^\dag_i(C) |0\ra
\eeq
All other states can be related to these states by certain permutations.

Now we calculate the matrix elements of the effective model between these states at the order of $t^3/U^2$. The matrix elements between different types of the above states always vanish. So we only need to calculate the matrix elements between states within the same type.

\beq
\la ij, ij, ij|H^{(3)}|ij, ij, ij\ra=0
\eeq

\beq
\begin{split}
\la ij, jk, ij|H^{(3)}|ij, jk, ij\ra&=-\frac{t^3}{U^2}(2\cos(3\eta_k\phi)-2\cos(3\eta_i\phi))=0\\
\la ij, ij, jk|H^{(3)}|ij, jk, ij\ra&=-\frac{3t^3}{U^2}\left(e^{i\eta_k\phi+2i\eta_i\phi} -e^{-2i\eta_k\phi-i\eta_i\phi}\right)
\end{split}
\eeq

\beq
\begin{split}
\la ij, jk, ik|H^{(3)}|ij, jk, ik\ra&=-\frac{t^3}{U^2}(-2\cos(3\eta_i\phi)-2\cos(3\eta_j\phi) -2\cos(3\eta_k\phi))=\frac{6t^3}{U^2}\cos 3\phi\\
\la ij, ik, jk|H^{(3)}|ij, jk, ik\ra&=-\frac{3t^3}{U^2}(e^{2i\eta_i\phi+i\eta_j\phi} +e^{-2i\eta_j\phi-i\eta_i\phi})
\end{split}
\eeq

\beq
\begin{split}
\la ij, kl, il|H^{(3)}|ij, kl, il\ra&=-\frac{t^3}{U^2}(2\cos(3\eta_j\phi)+2\cos(3\eta_k\phi) -2\cos(3\eta_l\phi)-2\cos(3\eta_i\phi))=0\\
\la ij, kl, il|H^{(3)}|ij, il, kl\ra&=\frac{3t^3}{U^2}(e^{2i\eta_k\phi+i\eta_i\phi} -e^{-i\eta_k\phi-2i\eta_i\phi}) \\
\la ij, kl, ij|H^{(3)}|ij, kl, ij\ra&=-\frac{t^3}{U^2}(2\cos(3\eta_k\phi)+2\cos(3\eta_l\phi) -2\cos(3\eta_j\phi)-2\cos(3\eta_i\phi))=0\\
\la ij, kl, ij|H^{(3)}|ij, ij, kl\ra&=0\\
\la ij, kl, ij|H^{(3)}|ik, lj, ij\ra&=-\frac{3t^3}{U^2}(e^{i\eta_k\phi+2i\eta_j\phi} -e^{-2i\eta_k\phi-i\eta_j\phi})
\end{split}
\eeq

\beq
\begin{split}
\la ij, ik, il|H^{(3)}|ij, ik, il\ra&=-\frac{t^3}{U^2}(2\cos(3\eta_j\phi)+2\cos(3\eta_k\phi) +2\cos(3\eta_l\phi))=-\frac{6t^3}{U^2}\cos3\phi\\
\la ij, il, ik|H^{(3)}|ij, ik, il\ra&=\frac{3t^3}{U^2}(e^{i\eta_k\phi+2i\eta_l\phi} +e^{-i\eta_l\phi-2i\eta_k\phi})\\
\la il, ij, ik|H^{(3)}|ij, ik, il\ra&=-\frac{6t^3}{U^2}e^{i\eta_j\phi+i\eta_k\phi+i\eta_l\phi}
\end{split}
\eeq

Recasting these matrix elements into a compact form yields
\beq
\begin{split}
H^{(3)}
=&-\frac{6t^3}{U^2}
\left[
T_A^{lj}T_B^{jk}T_C^{kl}e^{2i\phi(\tilde{I}_A^z+\tilde{I}_B^z+\tilde{I}_C^z)}
+e^{-2i\phi(\tilde{I}_A^z+\tilde{I}_B^z+\tilde{I}_C^z)}T_A^{jl}T_B^{kj}T_C^{lk}
\right]\\
&+\frac{3t^3}{U^2}
\left[
T_{B}^{ij}T_C^{ji}e^{2i\phi(\tilde{I}_B^z+2\tilde{I}_C^z)}+\hc
+(B\rightarrow C\rightarrow A\rightarrow B)
\right]
-\frac{12t^3}{U^2}\cos3\phi
\end{split}
\eeq
where $\tilde{I}_A^z$ gives the flavor of the particle that is acted by the $T$ operators.

\subsection{Effective models in terms of spin and orbital operators}

As seen in the above, the effective Hamiltonian expressed in terms the operators $T^{ij}$ is relatively concise, and they are the same for both $N_0=-1$ and $N_0=-2$ up to some constants. However, to gain more intuition, it is helpful to express these effective Hamiltonians in terms of spin operators $\vec S$ and valley operator $\vec I$, where
\beq
\begin{split}
&S^+=c^\dag_{+\uparrow}c_{+\downarrow}+c^\dag_{-\uparrow}c_{-\downarrow}=T^{12}+T^{34},
\quad
S^-=c^\dag_{+\downarrow}c_{+\uparrow}+c^\dag_{-\downarrow}c_{-\uparrow}=T^{21}+T^{43},\\
&S^z=\frac{1}{2}(c^\dag_{+\uparrow}c_{+\uparrow}+c^\dag_{-\uparrow}c_{-\uparrow} -c^\dag_{+\downarrow}c_{+\downarrow}-c^\dag_{-\downarrow}c_{-\downarrow})
=\frac{1}{2}(T^{11}+T^{33}-T^{22}-T^{44})\\
&I^+=c^\dag_{+\uparrow}c_{-\uparrow}+c^\dag_{+\downarrow}c_{-\downarrow}=T^{13}+T^{24},
\quad
I^-=c^\dag_{-\uparrow}c_{+\uparrow}+c^\dag_{-\downarrow}c_{+\downarrow}=T^{31}+T^{42},\\
&I^z=\frac{1}{2}(c^\dag_{+\uparrow}c_{+\uparrow}+c^\dag_{+\downarrow}c_{+\downarrow} -c^\dag_{-\uparrow}c_{-\uparrow}-c^\dag_{-\downarrow}c_{-\downarrow})
=\frac{1}{2}(T^{11}+T^{22}-T^{33}-T^{44})
\end{split}
\eeq
As well as filling fraction
\beq
n=\frac{1}{2}\left(c^\dag_{+\uparrow}c_{+\uparrow}+c^\dag_{+\downarrow}c_{+\downarrow} +c^\dag_{-\uparrow}c_{-\uparrow}+c^\dag_{-\downarrow}c_{-\downarrow}\right)
=\frac{1}{2}\left(T^{11}+T^{22}+T^{33}+T^{44}\right)
\eeq
Here $n=1/2$ means $N_0=1$ and $n=1$ means $N_0=2$. In the above $\vec S$ and $\vec I$ form two decoupled $SU(2)$ algebras, and $n$ commutes with all others.

\subsubsection{Effective Hamiltonian for $N_0=1$}

To this end, we first re-express the operators $T^{ij}$ in terms of $\vec S$, $\vec I$ and $n$. For $n=1/2$ ($N_0=1$),
\beq
\begin{split}
	T^{11}=(n+S^z)(n+I^z),\ T^{12}=S^+(n+I^z),\ T^{13}=I^+(n+S^z),\ T^{14}&=S^+I^+,\\
	             T^{22}=(n-S^z)(n+I^z),\ T^{23}=S^-I^+,\ T^{24}&=I^+(n-S^z),\\
	                           T^{33}=(n+S^z)(n-I^z),\ T^{34}&=S^+(n-I^z),\\
	                                         T^{44}&=(n-S^z)(n-I^z)
\end{split}
\eeq
Substituting these into (\ref{eq: 2nd order N0=1}) and (\ref{eq: 3rd order N0=1}) yields
\beq \label{eq: converting operators}
\begin{split}
&\left(H^{(2)}+\frac{2t^2}{U}\right)\cdot\frac{U}{2t^2}
=\sum_{ij}T_A^{ij}T_B^{ji}\\
=&\frac{1}{2}(S_A^+S_B^-+I_A^+I_B^-+\hc)+(S_A^+S_B^-+S_A^-S_B^+)(I_A^+I_B^-+I_A^-I_B^+)\\
&+2I_A^zI_B^z(S_A^+S_B^-+S_A^-S_B^+)+2S_A^zS_B^z(I_A^+I_B^-+I_A^-I_B^+)\\
&+4(n^2+S_A^zS_B^z)(n^2+I_A^zI_B^z)\\
=&4\cdot\left(n^2+\vec S_A\cdot\vec S_B\right)\cdot\left(n^2+\vec I_A\cdot\vec I_B\right)
\end{split}
\eeq
and
\beq
\begin{split}
	\left(H^{(3)}+\frac{6t^3}{U^2}\cos 3\phi\right)\frac{U^2}{3t^3}=-2h^{(3,1)}+h^{(3,2)}
\end{split}
\eeq
with
\beq \label{eq: h-31}
\begin{split}
	h^{(3,1)}
	=&8n^2\cos 3\phi(n^2+\vec S_A\cdot\vec S_B+\vec S_B\cdot\vec S_C+\vec S_C\cdot\vec S_A)(n^2+I_A^zI_B^z+I_B^zI_C^z+I_C^zI_A^z)\\
	+&4n^2(n^2+\vec S_A\cdot\vec S_B+\vec S_B\cdot\vec S_C+\vec S_C\cdot\vec S_A)\cdot\left[e^{i\phi}(I_A^+I_B^-+I_B^+I_C^-+I_C^+I_B^-)+\hc\right]\\
	+&8\sin 3\phi[I_A^zI_B^zI_C^z+n^2(I_A^z+I_B^z+I_C^z)]\cdot\left[\vec S_A\cdot(\vec S_B\times\vec S_C)\right]\\
	-&4i\vec S_A\cdot(\vec S_B\times\vec S_C)\cdot
	\left[
	I_A^z(e^{i\phi}I_B^+I_C^--e^{-i\phi}I_B^-I_C^+)+(A\rightarrow B\rightarrow C\rightarrow A)
	\right]
\end{split}
\eeq
and
\beq \label{eq: h-32}
\begin{split}
	h^{(3,2)}
	&=T_B^{kj}T_C^{jk}e^{2i\phi(I_B^z+2I_C^z)}
	+e^{-2i\phi(I_B^z+2I_C^z)}T_B^{jk}T_C^{kj}
	+(B\rightarrow C\rightarrow A\rightarrow B)\\
	&=4\cdot\left(n^2+\vec S_A\cdot\vec S_B\right)\cdot\left(2n^2\cos 3\phi+2I_A^zI_B^z\cos 3\phi+e^{-i\phi}I_A^+I_B^-+e^{i\phi} I_A^-I_B^+\right)+(B\rightarrow C\rightarrow A\rightarrow B)
\end{split}
\eeq

\subsubsection{Effective Hamiltonian for $N_0=2$}

For $n=1$ ($N_0=2$), it is useful to first consider the general relation $T^{ij}T^{kl}=\delta_{jk}T^{il}-c^\dag_i c^\dag_k c_j c_i$. Restrcting to 2-particle states, we can use this general relation to write down
\beq
\begin{split}
	&S^+I^+=2T^{14},
	\quad
	S^+I^-=2T^{32},
	\quad
	S^+I^z=T^{12}-T^{34},\\
	&S^-I^+=2T^{23},
	\quad
	S^-I^-=2T^{41},
	\quad
	S^-I^z=T^{21}-T^{43},\\
	&S^zI^+=T^{13}-T^{24},
	\quad
	S^zI^-=T^{31}-T^{42},
	\quad
	S^zI^z=\frac{1}{2}(T^{11}-T^{22}-T^{33}+T^{44})
\end{split}
\eeq
Using these, we can convert the relations and get
\beq
\begin{split}
	T^{11}=\frac{(n+S^z)(n+I^z)}{2},\ T^{12}=\frac{S^+(n+I^z)}{2},\ T^{13}=\frac{I^+(n+S^z)}{2},\ T^{14}&=\frac{S^+I^+}{2},\\
	T^{22}=\frac{(n-S^z)(n+I^z)}{2},\ T^{23}=\frac{S^-I^+}{2},\ T^{24}&=\frac{I^+(n-S^z)}{2},\\
	T^{33}=\frac{(n+S^z)(n-I^z)}{2},\ T^{34}&=\frac{S^+(n-I^z)}{2},\\
	T^{44}&=\frac{(n-S^z)(n-I^z)}{2}
\end{split}
\eeq
which differs from (\ref{eq: converting operators}) only by factors of 2.

Substituting these into (\ref{eq: 2nd order N0=2}) and (\ref{eq: 3rd order N0=2}) yields
\beq
\left(H^{(2)}+\frac{4t^2}{U}\right)\frac{U}{2t^2}=\sum_{ij}T_A^{ij}T_B^{ji}=\left(n^2+\vec S_A\cdot\vec S_B\right)\cdot\left(n^2+\vec I_A\cdot\vec I_B\right)
\eeq
and
\beq
\left(H^{(3)}+\frac{12t^3}{U^2}\cos 3\phi\right)\cdot\frac{U^2}{3t^3}=-2h^{(3,1)}+h^{(3,2)}
\eeq
with $8h^{(3,1)}$ and $4h^{(3,2)}$ given by the same expressions as in (\ref{eq: h-31}) and (\ref{eq: h-32}), respectively, but notice the value of $n$ is changed from $1/2$ to $1$.

\end{document}